\documentclass[10pt,english,two column]{IEEEtran}
\usepackage[T1]{fontenc}
\usepackage[latin9]{inputenc}
\usepackage{color}
\usepackage{array}
\usepackage{float}
\usepackage{multirow}
\usepackage{amsmath}
\usepackage{amssymb}
\usepackage{graphicx}
\usepackage{esint}

\makeatletter

\providecommand{\tabularnewline}{\\}
\floatstyle{ruled}
\newfloat{algorithm}{tbp}{loa}
\providecommand{\algorithmname}{Algorithm}
\floatname{algorithm}{\protect\algorithmname}

\ifCLASSINFOpdf
\else
\fi
\usepackage{babel}

\usepackage{eso-pic}

\AddToShipoutPicture{%
  \AtPageUpperLeft{%
    \hspace*{20pt}\makebox(200,-20)[lt]{%
      \footnotesize%
      \textbf{TO APPEAR IN IEEE TRANSACTIONS ON VEHICULAR TECHNOLOGY, 2016}%
}}}

\makeatother

\usepackage{babel}
\begin{document}

\title{Modulation Classification for MIMO-OFDM Signals via Approximate Bayesian
Inference }

\author{Yu Liu, Osvaldo Simeone,\IEEEmembership{\ Fellow, IEEE,} Alexander
M. Haimovich,~\IEEEmembership{Fellow, IEEE,} Wei Su,~\IEEEmembership{Fellow, IEEE}
\thanks{Copyright (c) 2015 IEEE. Personal use of this material is permitted.
However, permission to use this material for any other purposes must
be obtained from the IEEE by sending a request to pubs-permissions@ieee.org. 

Y. Liu, O. Simeone and A. M. Haimovich are with the Center for Wireless
Communications and Signal Processing Research (CWCSPR), ECE Department,
New Jersey Institute of Technology (NJIT), Newark, NJ 07102, USA (email:
\{yl227, osvaldo.simeone, haimovic\}@njit.edu). 

W. Su is with the U.S. Army Communication-Electronics Research Development
and Engineering Center, I2WD, Aberdeen Proving Ground, MD 21005, USA
(email: wei.su@ieee.org).

DISTRIBUTION STATEMENT A: Approved for public release, distribution
is unlimited}}
\maketitle
\begin{abstract}
The problem of modulation classification for a multiple-antenna (MIMO)
system employing orthogonal frequency division multiplexing (OFDM)
is investigated under the assumption of unknown frequency-selective
fading channels and signal-to-noise ratio (SNR). The classification
problem is formulated as a Bayesian inference task, and solutions
are proposed based on Gibbs sampling and mean field variational inference.
The proposed methods rely on a selection of the prior distributions
that adopts a latent Dirichlet model for the modulation type and on
the Bayesian network formalism. The Gibbs sampling method converges
to the optimal Bayesian solution and, using numerical results, its
accuracy is seen to improve for small sample sizes when switching
to the mean field variational inference technique after a number of
iterations. The speed of convergence is shown to improve via annealing
and random restarts. While most of the literature on modulation classification
assume that the channels are flat fading, that the number of receive
antennas is no less than that of transmit antennas, and that a large
number of observed data symbols are available, the proposed methods
perform well under more general conditions. Finally, the proposed
Bayesian methods are demonstrated to improve over existing non-Bayesian
approaches based on independent component analysis and on prior Bayesian
methods based on the `superconstellation' method. \end{abstract}


\sloppy

\section{Introduction}

Cognitive radio is a wireless communication technology that addresses
the inefficiency of the radio resource usage via computational intelligence
\cite{Mitola,Dobre-1}. Cognitive radios have both civilian and military
applications \cite{Dobre}. A major task in such radios is the classification
of the modulation format of unknown received signals. As the pairing
of multiple-antenna (MIMO) transmission and orthogonal frequency division
multiplexing (OFDM) data modulation has become central to fourth generation
(4G) and fifth generation (5G) wireless technologies, a need has arisen
for the development of new classification algorithms capable of handling
MIMO-OFDM signals. 

Modulation classification methods are generally classified as inference-based
or pattern recognition-based \cite{Dobre}. The inference-based approaches
fall into two categories, namely Bayesian and non-Bayesian methods
\cite{Gelman}. Bayesian methods model unknown parameters as random
variables with some prior distributions, and aim to evaluate the posterior
probability of the modulation type. Non-Bayesian methods, instead,
model unknown parameters as nuisance variables that need to be estimated
before performing modulation classification. With pattern recognition-based
methods, specific features are extracted from the received signal
and then used to discriminate among the candidate modulations. Compared
to the pattern recognition-based approaches, inference-based methods
generally achieve better classification performance at the cost of
a higher computational complexity \cite{Dobre}.

There is ample literature on classification algorithms for single-antenna
(SISO) systems \cite{Dobre}, \cite{Hameed}-\cite{Y. Liu}. Among
them, a Bayesian method using Gibbs sampling is proposed in \cite{Thomas}.
In \cite{Y. Liu}, a systematic Bayesian solution based on the latent
Dirichlet Bayesian Network (BN) is proposed, which generalizes and
improves upon the work in \cite{Thomas}. A preprocessor for modulation
classification is developed in \cite{Amuru}, whereby the timing offset
is estimated using grid-based Gibbs sampling\footnote{In grid-based Gibbs sampling, a grid of points is first selected within
the domain of a variable $x$ to be sampled. Then the conditional
probability density function (normalized or non-normalized) of variable
$x$ is computed at each selected point, which is used for sampling
$x$.}. We note that, while most algorithms rely on the assumption that
the channels are flat fading or additive white Gaussian noise channels,
the approaches in \cite{Thomas}-\cite{Amuru} are well-suited also
for frequency selective fading channels. 

Only few publications address modulation classification in MIMO systems
\cite{Choqueuse}-\cite{Marey}. The task of modulation classification
for MIMO is more challenging than for SISO due to the mutual interference
between the received signals and to the multiplicity of unknown channels.
In \cite{Choqueuse}, a non-Bayesian approach, referred to as independent
component analysis (ICA)-phase correction (PC), is proposed, where
the channel matrix required for the calculation of the hypotheses
test is estimated blindly by ICA \cite{Hyvarinen}. Several related
pattern recognition-based algorithms are introduced in \cite{Muhlhaus}-\cite{Hassan},
where source streams are separated by ICA-PC or a constant modulus
algorithm, and diverse higher-order signal statistics are used as
discriminating features. Moreover, a pattern recognition-based algorithm
using spatial and temporal correlation functions as distinctive features
is reported in \cite{Marey} for MIMO frequency-selective channels
with time-domain transmission. The approach is not applicable to modulation
classification for MIMO-OFDM systems. As for MIMO-OFDM systems, a
non-Bayesian approach is proposed in \cite{Handan} based on ICA-PC
and an assumed invariance of the frequency-domain channels across
the coherence bandwidth. It is finally noted that the results in this
paper were partially presented in \cite{Liu_CISS}.

\textit{Main Contributions:} In this work, we develop Bayesian modulation
classification techniques for MIMO-OFDM systems operating over frequency-selective
fading channels, assuming unknown channels and signal-to-noise ratio
(SNR). Our main contributions are as follows.
\begin{enumerate}
\item A modulation classification technique is proposed based on Gibbs sampling
for MIMO-OFDM systems. Inspired by the latent Dirichlet models in
machine learning \cite{Blei-1}, this approach leverages a novel selection
of the prior distributions for the unknown variables, the modulation
format and the transmitted symbols. This selection was first adopted
by some of the authors in \cite{Y. Liu} for SISO systems. As compared
to SISO systems, a Gibbs sampling implementation such as in \cite{Y. Liu}
may have an impractically slow convergence due to the high-dimensional
and multimodal distributions in MIMO systems. The strategy of annealing
\cite{Kirkpatrick}-\cite{Fevotte} combined with multiple random
restarts \cite{Chockalingam}-\cite{Wang} is hence proposed here
to improve the convergence speed. 
\item An alternative Bayesian solution for modulation classification in
MIMO-OFDM systems that leverages mean field variational inference
\cite{Koller} is proposed, based on the same latent Dirichlet prior
selection.
\item A hybrid approach that switches from Gibbs sampling to mean field
variational inference is proposed for modulation classification in
MIMO-OFDM systems. The hybrid approach is motivated by the fact that
the Gibbs sampler is superior to mean field as a method for exploring
the global solution space, while the mean field algorithm has better
convergence speed in the vicinity of a local optima \cite{Koller}-\cite{Blei}. 
\item Extensive numerical results demonstrate that the proposed Gibbs sampling
method converges to an effective solution, and its accuracy improves
for small sample sizes when switching to the mean field variational
inference technique after a number of iterations. Moreover, the speed
of convergence is seen to be generally improved by multiple random
restarts and annealing \cite{Kirkpatrick}-\cite{Wang}. Overall,
while most of the reviewed existing modulation classification algorithms
for MIMO-OFDM systems work under the assumptions that the channels
are flat fading \cite{Choqueuse}-\cite{Hassan}, that the number
of receive antennas is no less than the number of transmit antennas
\cite{Choqueuse}-\cite{Hassan}, and/or that the number of samples
is large (as for pattern recognition-based methods) \cite{Muhlhaus}-\cite{Marey},
the proposed method achieves satisfactory performance under more general
conditions.
\end{enumerate}
The rest of the paper is organized as follows. The signal model is
introduced in Sec. \ref{sec:System-Model}. In Sec. \ref{sec:Preliminaries},
we briefly review some necessary preliminary concepts, including Bayesian
inference and BNs, while in Sec. \ref{sec:Bayesian-Inference-for},
we formulate the modulation classification problem under study as
a Bayesian inference task, and propose solutions based on Gibbs sampling
and on mean field variational inference. Numerical results of the
proposed methods are presented in Sec. \ref{sec:Numerical-results-and.}.
Finally, conclusions are drawn in Sec. \ref{sec:Conclusions}. 

\textit{Notation}: The superscripts $T$ and $H$ denote matrix or
vector transpose and Hermitian, respectively. The $i$-th row of the
matrix $\mathbf{B}$ is denoted as $\mathbf{B}_{(i,\cdot)}$ and the
$j$-th column is denoted as $\mathbf{B}_{(\cdot,j)}$. Lower case
bold letters and upper case bold letters are used to denote column
vectors and matrices, respectively. The notation $\mathbf{b}\diagdown b_{i}$,
where $\mathbf{b}=\left[b_{1},...,b_{n}\right]^{T}$ and $i\in\{1,...,n\}$,
denotes the vector composed of all the elements of $\mathbf{b}$ except
$b_{i}$. We use an angle bracket $\langle\cdot\rangle_{\centerdot}$
to represent the expectation with respect to the random variables
indicated in the subscript. For notational simplicity, we do not indicate
the variables in the subscript when the expectation is taken with
respect to all the random variables inside the bracket $\langle\cdot\rangle_{\centerdot}$
The notations $\psi(\cdot)$ and $\mathrm{\mathbf{1}}(\cdot)$ stand
for the digamma function \cite{Abramowitz} and the indicator function,
respectively. The cardinality of a set $\mathcal{B}$ is denoted $\left|\mathcal{B}\right|$.
We use the same notation, $p(\cdot)$, for both probability density
functions (pdf) and probability mass function (pmf). Moreover, we
will identify a pdf or pmf by its arguments, e.g., $p(X|Y)$ represents
the distribution of random variable $X$ given the random variable
$Y$. The notations $\mathcal{CN}(\mathbf{\boldsymbol{\mu}},\mathbf{C})$
and $\mathcal{IG}\left(a,b\right)$ represent the the circularly symmetric
complex Gaussian distribution with mean vector $\boldsymbol{\mu}$
and covariance matrix $\mathbf{C}$ and the inverse gamma distribution
with shape parameter $a$ and scale parameter $b$, respectively.

\section{System Model\label{sec:System-Model}}

Consider a MIMO-OFDM system operating over a frequency-selective fading
channel with $N$ subcarriers, $M_{t}$ transmit antennas, $M_{r}$
receive antennas and a coherence period of $K$ OFDM symbols. All
frequency-domain symbols transmitted during the coherence period are
taken from a finite constellation $A\in\mathcal{A}$, such as $M$-PSK
or $M$-QAM, where $\mathcal{A}$ is the (finite) set containing all
possible constellations. Without loss of generality, $A$ is assumed
to be a constellation of symbols with average power equal to 1. The
number of transmit antennas $M_{t}$ is assumed known. We observe
that several algorithms have been proposed for the estimation of $M_{t}$
\cite{Aouda,Somekh}. We focus on the problem of detecting the constellation
$A$ in the absence of information about the SNR, the transmitted
symbols and the fading channel coefficients. 

After matched filtering and sampling, assuming that time synchronization
has been successfully performed at least within the error margin afforded
by the cyclic prefix, the frequency-domain samples $\mathbf{y}[n,k]=[y_{1}[n,k],...,y_{M_{r}}[n,k]]^{T}$,
received across the $M_{r}$ receive antennas, at the $n$-th subcarrier
of the $k$-th OFDM frame, are expressed as
\begin{equation}
\mathbf{y}[n,k]=\mathbf{H}[n]\mathbf{s}[n,k]+\mathbf{z}[n,k],\label{eq:instantaneous mixtures}
\end{equation}
where $\mathbf{H}[n]$ is the $M_{r}\times M_{t}$ frequency-domain
channel matrix associated with the $n$-th subcarrier; $\mathbf{s}[n,k]$
is the $M_{t}\times1$ vector composed of the symbols transmitted
by the $M_{t}$ antennas, i.e., $\mathbf{s}[n,k]=[s_{1}[n,k],...,s_{M_{t}}[n,k]]^{T}$,
with $s_{m_{t}}[n,k]\in A$ being the symbol transmitted by the $m_{t}$-th
transmit antenna over the $n$-th subcarrier of the $k$-th OFDM symbol;
and $\mathbf{z}[n,k]=[z_{1}[n,k],...,z_{M_{r}}[n,k]]^{T}\sim\mathcal{CN}(\mathbf{0},\sigma^{2}\mathbf{I})$
is complex white Gaussian noise, which is independent over indices
$n$ and $k$. The frequency-domain channel matrix $\mathbf{H}[n]$
can be written 
\begin{equation}
\mathbf{H}[n]=\left[\begin{array}{ccc}
\tilde{h}_{1,1}\left[n\right] & \cdots & \tilde{h}_{M_{t},1}\left[n\right]\\
\vdots & \ddots & \vdots\\
\tilde{h}_{1,M_{r}}\left[n\right] & \cdots & \tilde{h}_{M_{t},M_{r}}\left[n\right]
\end{array}\right],\label{eq:H_n}
\end{equation}
where $\tilde{\mathbf{h}}_{m_{t},m_{r}}=[\tilde{h}_{m_{t},m_{r}}\left[1\right],...,\tilde{h}_{m_{t},m_{r}}\left[N\right]]^{T}$
denotes the $N\times1$ frequency-domain channel vector between the
$m_{t}$-th transmit antenna and the $m_{r}$-th receive antenna.
Assuming that the channel for any pair $(m_{t},m_{r})$ has at most
$L$ symbol-spaced taps, we write $\tilde{\mathbf{h}}_{m_{t},m_{r}}=\mathbf{\mathbf{W}}\mathbf{h}_{m_{t},m_{r}}$,
with $\boldsymbol{\mathbf{h}}_{m_{t},m_{r}}$ the $L\times1$ time-domain
channel vector and $\mathbf{W}$ the $N\times L$ matrix composed
of the first $L$ columns of the DFT matrix of size $N$. Note that
the channel is fixed within the coherence frame of $K$ OFDM symbols. 

According to (\ref{eq:instantaneous mixtures}) and (\ref{eq:H_n}),
the $NK\times1$ received frequency-domain signals $\mathbf{y}_{m_{r}}=[\mathbf{y}_{m_{r}}[1]^{T},...,$$\mathbf{y}_{m_{r}}[K]^{T}]^{T}$
at the $m_{r}$-th receive antenna is given by
\begin{equation}
\mathbf{y}_{m_{r}}=\sum_{m_{t}=1}^{M_{t}}\mathbf{D}_{m_{t}}\tilde{\mathbf{h}}_{m_{t},m_{r}}+\mathbf{z}_{m_{r}},\:m_{r}=1,...,M_{r},\label{eq:sig_mod_1}
\end{equation}
where $\mathbf{y}_{m_{r}}[k]=[y_{m_{r}}[1,k],...,y_{m_{r}}[N,k]]^{T}$;
$\mathbf{D}_{m_{t}}=[\mathbf{D}_{m_{t},1},...,\mathbf{D}_{m_{t},K}]^{T}$
is an $NK\times N$ matrix representing the transmitted symbols with
$\mathbf{D}_{m_{t},k}$ an $N\times N$ diagonal matrix whose $(n,n)$
element is $s_{m_{t}}[n,k]$; and $\mathbf{z}_{m_{r}}=[\mathbf{z}_{m_{r}}[1]^{T},...,\mathbf{z}_{m_{r}}[K]^{T}]^{T}$
with $\mathbf{z}_{m_{r}}[k]=[z_{m_{r}}[1,k],...,z_{m_{r}}[N,k]]^{T}$.

Let us further define the $NKM_{t}\times1$ vector $\mathbf{s}=[\mathbf{s}_{1},..,\mathbf{s}_{K}]^{T}$
containing all the transmitted symbols with $\mathbf{s}_{k}=[\mathbf{s}[1,k]^{T},...,\mathbf{s}[N,k]^{T}]^{T}$;
the $LM_{t}M_{r}\times1$ vector $\mathbf{h}=[\mathbf{h}_{1}^{T},...,\mathbf{h}_{M_{r}}^{T}]^{T}$
for the time domain channels associated with all the transmit-receive
antenna pairs, where $\mathbf{h}_{m_{r}}=[\boldsymbol{\mathbf{h}}_{1,m_{r}}^{T},...,\boldsymbol{\mathbf{h}}_{M_{t},m_{r}}^{T}]^{T}$;
and the $NKM_{r}\times1$ receive signal vector $\mathbf{y}=[\mathbf{y}_{1}^{T},...,\mathbf{y}_{M_{r}}^{T}]^{T}$.
The task of modulation classification is for the receiver to correctly
detect the modulation format $A$ given only the received samples
$\mathbf{y}$, while being uninformed about the symbols $\mathbf{s}$,
the channel $\mathbf{h}$ and the noise power $\sigma^{2}$. Using
(\ref{eq:instantaneous mixtures}) and (\ref{eq:sig_mod_1}), the
likelihood function $p\left(\mathbf{y}|A,\mathbf{s},\mathbf{h}\mathrm{,\sigma^{2}}\right)$
of the observation is given by
\begin{align}
 & p\left(\mathbf{y}\Big|A,\mathbf{s},\mathbf{h}\mathrm{,\sigma^{2}}\right)\nonumber \\
= & \prod_{n,k}p\left(\mathbf{y}[n,k]\Big|\mathbf{s}[n,k],\mathbf{H}[n],\sigma^{2}\right)\nonumber \\
= & \prod_{m_{r}}p\left(\mathbf{y}_{m_{r}}\Big|\mathbf{s},\mathbf{h}_{m_{r}}\mathrm{,\sigma^{2}}\right),\label{eq:joint Likelihood-1}
\end{align}
with $\mathbf{y}_{m_{r}}|(\mathbf{s},\mathbf{h}_{m_{r}}\mathrm{,\sigma^{2}})$
$\sim$ $\mathcal{CN}(\sum_{m_{t=1}}^{M_{t}}\mathbf{D}_{m_{t}}\mathbf{\mathbf{W}}\mathbf{h}_{m_{t},m_{r}},\sigma^{2}\mathbf{I})$
and $\mathbf{y}[n,k]|(\mathbf{s}[n,k],\mathbf{H}[n],\sigma^{2})$
$\sim$$\mathcal{CN}(\mathbf{H}[n]\mathbf{s}[n,k],\sigma^{2}\mathbf{I})$.

\section{Preliminaries\label{sec:Preliminaries} }

As formalized in the next section, in this paper we formulate the
modulation classification problem as a Bayesian inference task. In
this section, we review some necessary preliminary concepts. Specifically,
we start by introducing the general task of Bayesian inference in
Sec. \ref{sub:Bayesian-Inference}; we review the definition of BN,
which is a useful graphical tool to represent knowledge about the
structure of a joint distribution, in Sec. \ref{sub:Bayesian-Network};
and, finally, we review approximate solutions to the Bayesian inference
task, namely, Gibbs sampling in Sec. \ref{sub:Gibbs-sampling}, and
mean field variational inference in Sec. \ref{sub:Mean-Field-Variational}.

\subsection{Bayesian Inference\label{sub:Bayesian-Inference}}

Bayesian inference aims to compute the posterior probability of the
variables of interest given the evidence, where the evidence is a
subset of the random variables in the model. Specifically, given the
values of some evidence variables $\mathbf{\Theta}_{e}=\mathbf{\mathbf{\boldsymbol{\theta}}}_{e}$,
one wishes to estimate the posterior distribution of a subset of the
unknown variables $\mathbf{\Theta}_{u}=[\Theta_{1},...,\Theta_{G}]^{T}$.
We assume here for simplicity of exposition that all variables are
discrete with finite cardinality. However, the extension to continuous
variables with pdfs is immediate, as it will be argued. The conditional
pmf of $\mathbf{\Theta}_{u}$ given the evidence $\mathbf{\Theta}_{e}=\mathbf{\mathbf{\boldsymbol{\theta}}}_{e}$
is proportional to the product of a prior distribution $p(\mathbf{\Theta}_{u})$
on the unknown variables $\mathbf{\Theta}_{u}$ and of the likelihood
of the evidence $p(\mathbf{\Theta}_{e}|\mathbf{\Theta}_{u})$:
\begin{equation}
p(\mathbf{\Theta}_{u}|\mathbf{\Theta}_{e}=\mathbf{\boldsymbol{\theta}}_{e})\propto p(\mathbf{\Theta}_{u})p(\mathbf{\Theta}_{e}=\mathbf{\mathbf{\boldsymbol{\theta}}}_{e}|\mathbf{\Theta}_{u}).\label{eq:Bayes' theorem}
\end{equation}
If one is interested in computing the posterior distribution of the
unknown variable $\Theta_{j}$, then a direct approach would be to
write
\begin{equation}
p(\Theta_{j}=\mathbf{\theta}_{j}|\mathbf{\Theta}_{e}=\mathbf{\boldsymbol{\theta}}_{e})=\sum_{\mathbf{\boldsymbol{\theta}}_{u}\diagdown\mathbf{\theta}_{j}}p(\mathbf{\Theta}_{u}=\mathbf{\boldsymbol{\theta}}_{u}|\mathbf{\Theta}_{e}=\mathbf{\boldsymbol{\theta}}_{e}).\label{eq:general Marginal}
\end{equation}
The inference task (\ref{eq:general Marginal}) is made difficult
in practice by the multidimensional summation over all the values
of the variables $\mathbf{\Theta}_{u}\diagdown\Theta_{j}$. Note also
that, if the variables are continuous, the operation of summation
is replaced by integration and a similar discussion applies. Next,
we discuss the BN model.

\subsection{Bayesian Network\label{sub:Bayesian-Network}}

A BN is an acyclic graph that can be used to represent useful aspects
of the structure of a joint distribution. Each node in the graph represents
a random variable, while the directed edges between the nodes encode
the probabilistic influence of one variable on another. Node $\Theta_{i}$
is defined to be a parent of $\Theta_{j}$, if an edge from node $\Theta_{i}$
to node $\Theta_{j}$ exists in the graph. According to the BN's chain
rule \cite{Koller}, the influence encoded in a BN for a set of variables
$\mathbf{\Theta}=[\Theta_{1},...,\Theta_{J}]^{T}$ can be interpreted
as the factorization of the joint distribution in the form
\begin{equation}
p\left(\mathbf{\Theta}\right)=\prod_{j=1}^{J}p\left(\Theta_{j}|\mathrm{P}\mathrm{a}_{\Theta_{j}}\right),\label{eq:BN's chain rule}
\end{equation}
where we use $\mathrm{P}\mathrm{a}_{\Theta_{j}}$ to denote the set
of parent variables of variable $\Theta_{j}$. Note that (\ref{eq:BN's chain rule})
encodes the fact that each variable $\Theta_{j}$ is independent of
its ancestors in the BN, when conditioning on its parent variables
$\mathrm{P}\mathrm{a}_{\Theta_{j}}$. In the following, we will find
it useful to rewrite (\ref{eq:BN's chain rule}) in a more abstract
way as \cite{Koller}
\begin{equation}
p\left(\mathbf{\Theta}\right)=\prod_{\phi}\phi\left(\mathcal{B}_{\phi}\right),\label{eq:newFactorization}
\end{equation}
where the product is taken over all $J$ factor $\phi(\mathcal{B}_{\phi})=p(\Theta_{j}|\mathrm{P}\mathrm{a}_{\Theta_{j}})$
with $\mathcal{B}_{\phi}=\{\Theta_{j},\mathrm{P}\mathrm{a}_{\Theta_{j}}\}$.

\subsection{Gibbs Sampling\label{sub:Gibbs-sampling}}

Markov chain Monte Carlo (MCMC) techniques provide effective iterative
approximate solutions to the Bayesian inference task (\ref{eq:general Marginal})
that are based on randomization and can obtain increasingly accurate
posterior distributions as the number of iterations increases. The
goal of these techniques is to generate $M$ random samples $\mathbf{\boldsymbol{\theta}}_{u}^{\left(1\right)},...,\mathbf{\mathbf{\boldsymbol{\theta}}}_{u}^{\left(M\right)}$
from the desired posterior distribution $p(\mathbf{\Theta}_{u}|\mathbf{\Theta}_{e}=\mathbf{\boldsymbol{\theta}}_{e})$.
This is done by running a Markov chain whose equilibrium distribution
is $p(\mathbf{\Theta}_{u}|\mathbf{\Theta}_{e}=\mathbf{\boldsymbol{\theta}}_{e})$.
As a result, according to the law of large numbers, the multidimensional
summation (or integration) (\ref{eq:general Marginal}) can be approximated
by an ensemble average. In particular, the marginal distribution of
any particular variable $\Theta_{j}$ in $\mathbf{\Theta}_{u}$ can
be estimated as
\begin{equation}
p\left(\Theta_{j}=\theta_{j}|\mathbf{\Theta}_{e}=\mathbf{\boldsymbol{\theta}}_{e}\right)\approx\frac{1}{M}\sum_{m=M_{0}+1}^{M}\mathrm{\mathbf{1}}\left(\mathbf{\theta}_{j}^{\left(m\right)}=\theta_{j}\right),\label{eq:empiricalAverage}
\end{equation}
where $\mathbf{\theta}_{j}^{\left(m\right)}$ is the $m$-th sample
for $\Theta_{j}$ generated by the Markov chain, and $M_{0}$ denotes
the number of samples used as burn-in period to reduce the correlations
with the initial values \cite{Robert}. 

Gibbs sampling is a classical MCMC algorithm that defines the aforementioned
Markov chain by sampling all the variables in $\mathbf{\Theta}_{u}$
one-by-one. Specifically, the algorithm begins with a set of arbitrary
feasible values for $\mathbf{\Theta}_{u}$. Then, at step $m$, a
sample for a given variable $\Theta_{j}$ is drawn from the conditional
distribution $p(\Theta_{j}|\mathbf{\Theta}_{u}\diagdown\Theta_{j},\mathbf{\Theta}_{e})$.
Whenever a sample is generated for a variable, the value of that variable
is updated within the vector $\mathbf{\Theta}_{u}$. It can be shown
that the required conditional distributions $p(\Theta_{j}|\mathbf{\Theta}_{u}\diagdown\Theta_{j},\mathbf{\Theta}_{e})$
may be calculated by multiplying all the factors in the factorization
(\ref{eq:newFactorization}) that contain the variable of interest
and then normalizing the resulting distribution, i.e., we have
\begin{equation}
p(\Theta_{j}|\mathbf{\Theta}_{u}\diagdown\Theta_{j},\mathbf{\Theta}_{e})\propto\prod_{\phi:\,\Theta_{j}\in B_{\phi}}\phi\left(\mathcal{B}_{\phi}\right),\label{eq:FullCondionalEq}
\end{equation}
where the right-hand side of (\ref{eq:FullCondionalEq}) is the product
of the factors in (\ref{eq:newFactorization}) that involve the variable
$\Theta_{j}$. 

\textit{Remark 1}: In order for Markov chain Monte Carlo algorithms
to converge to a unique equilibrium distribution, the associated Markov
chain needs to be irreducible and aperiodic \cite[Ch. 12]{Koller}.
In the context of the Gibbs sampling, a sufficient condition for asymptotic
correctness of Gibbs sampling is that the distributions $p(\Theta_{j}|\mathbf{\Theta}_{u}\diagdown\Theta_{j},\mathbf{\Theta}_{e})$
are strictly positive in their domains for all $j$.

\textit{Remark 2}: When applying Gibbs sampling to practical problems,
in particular those with high-dimensional and multimodal posterior
distribution $p(\Theta_{j}|\mathbf{\Theta}_{u}\diagdown\Theta_{j},\mathbf{\Theta}_{e})$,
slow convergence may be encountered due to the local nature of the
updates. One approach to address this issue is to run Gibbs sampling
with \textit{multiple random restarts} that are initialized with different
feasible solutions \cite{Chockalingam}-\cite{Wang}. Moreover, within
each run, \textit{simulated annealing} may be used to avoid low-probability
``traps.'' Accordingly, the prior probability, or the likelihood,
may be parametrized by a temperature parameter $T$, such that a large
temperature implies a lower reliance on the evidence aimed at exploring
more thoroughly the range of the variables. Samples are generated,
starting with a high temperature and ending with a low temperature
\cite{Kirkpatrick}-\cite{Fevotte}.

\subsection{Mean Field Variational Inference\label{sub:Mean-Field-Variational}}

Mean field variational inference provides an alternative way to approach
the Bayesian inference problem of calculating $p(\Theta_{j}=\mathbf{\theta}_{j}|\mathbf{\Theta}_{e}=\mathbf{\boldsymbol{\theta}}_{e})$.
The key idea of this method is that of searching for a distribution
$q(\mathbf{\Theta}_{u})$ that is closest to the desired posterior
distribution $p(\mathbf{\Theta}_{u}|\mathbf{\boldsymbol{\theta}}_{e})$,
in terms of the Kullback-Leibler (KL) divergence $\mathrm{KL}\left(q(\mathbf{\Theta}_{u})||p(\mathbf{\Theta}_{u}|\mathbf{\boldsymbol{\theta}}_{e})\right)$,
within the class $\mathcal{Q}$ of distributions that factorize as
the product of marginals, i.e., $q(\mathbf{\Theta}_{u})=\prod_{j=1}^{G}q(\Theta_{j})$
\cite[Ch. 11]{Koller}. The corresponding variational problem is given
as
\begin{align}
\mathrm{\underset{\mathit{q}}{minimize}} & \,\mathrm{KL}\left(q(\mathbf{\Theta}_{u})||p(\mathbf{\Theta}_{u}|\mathbf{\boldsymbol{\theta}}_{e})\right)\label{eq:MF_optimi}\\
 & \mathrm{s.t.}\;q\in\mathcal{Q}.\nonumber 
\end{align}
By imposing the necessary optimality conditions for problem (\ref{eq:MF_optimi}),
one can prove that the mean field approximation $q(\mathbf{\Theta}_{u})$
is locally optimal only if the proportionality \cite{Koller}
\begin{equation}
q\left(\Theta_{j}\right)\propto\exp\bigg\{\sum_{\phi:\,\Theta_{j}\in\mathcal{B}_{\phi}}\big\langle\ln\phi\left(\mathcal{B}_{\phi}\right)\big\rangle_{q\left(\mathcal{B}_{\phi}\diagdown\Theta_{j}\right)}\bigg\}\label{eq:fixed_point}
\end{equation}
holds for all $j=1,..,G$, where the expectation in (\ref{eq:fixed_point})
is taken with respect to the distribution $q(\mathcal{B}_{\phi}\diagdown\Theta_{j})=q(\mathbf{\Theta}_{u})/q(\Theta_{j})$.
The idea of the mean field variational inference is to solve (\ref{eq:fixed_point})
by means of fixed-point iterations (see \cite{Koller} for details).
It can be shown that each iteration of (\ref{eq:fixed_point}) results
in a better approximation $q$ to the target distribution $p(\mathbf{\Theta}_{u}|\mathbf{\boldsymbol{\theta}}_{e})$,
hence guaranteeing convergence to a local optimum of problem (\ref{eq:MF_optimi})
\cite[Sec. 11.5.1]{Koller}. Once an approximating distribution $q(\mathbf{\Theta}_{u})$
is obtained, an approximation of the desired posterior $p(\mathbf{\Theta}_{u}|\mathbf{\boldsymbol{\theta}}_{e})$
can be obtained as $p(\mathbf{\Theta}_{u}|\mathbf{\boldsymbol{\theta}}_{e})\approx q(\mathbf{\Theta}_{u}),$
and the marginal posterior distribution may be approximated as $p(\Theta_{j}|\mathbf{\boldsymbol{\theta}}_{e})\approx q(\Theta_{j}).$

\section{Bayesian Inference for Modulation Classification\label{sec:Bayesian-Inference-for}}

In this section, we solve the problem of detecting the modulation
$A\in\mathcal{A}$ by adopting a Bayesian inference formulation. First,
in Sec. \ref{sub:Modulation-Classification-via}, we discuss the problem
of selecting a proper prior distribution, and argue that a latent
Dirichlet model inspired by \cite{Blei-1} and first used for modulation
classification in \cite{Y. Liu}, provides an effective choice. Then,
based on this prior model, we develop two solutions, one based on
Gibbs sampling, in Sec. \ref{sub:Modulation-Classification-via-1},
and the other based on mean field variational inference, in Sec. \ref{sub:Modulation-classification-via}.

\subsection{Latent Dirichlet Bayesian Network\label{sub:Modulation-Classification-via}}

\begin{figure}[htbp]
\begin{centering}
\textsf{\includegraphics[width=7.5cm,height=5.5cm]{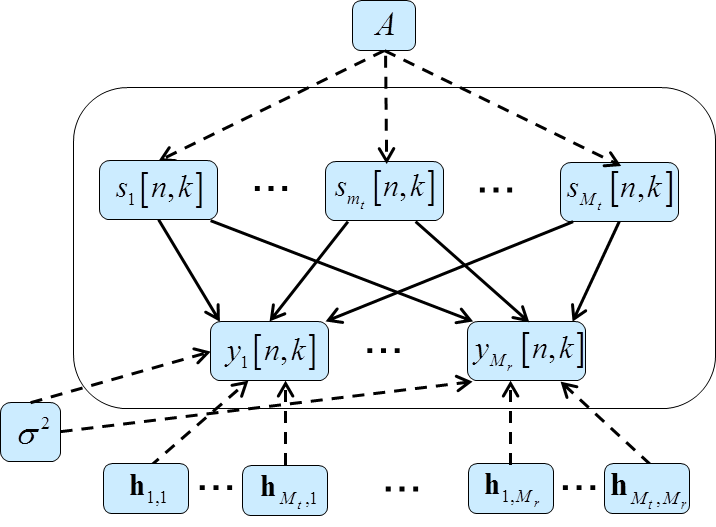}}
\par\end{centering}

\caption{\label{fig:Dirr_BN-1}BN \textsl{$\mathcal{G}_{1}$} for the modulation
classification scheme based on the factorization (\ref{eq:posterior_1}).
The nodes inside the rectangle are repeated $NK$ times.}
\end{figure}
According to (\ref{eq:Bayes' theorem}), the joint distribution of
the unknown variables $(A,\mathbf{s},\mathbf{h},\sigma^{2})$ may
be expressed
\begin{equation}
p\left(A,\mathbf{s},\mathbf{h}\mathrm{,\sigma^{2}}\Big|\mathbf{y}\right)\propto p\left(\mathbf{y}\Big|A,\mathbf{s},\mathbf{h}\mathrm{,\sigma^{2}}\right)p\left(A,\mathbf{s},\mathbf{h}\mathrm{,\sigma^{2}}\right),\label{eq:posterior_1}
\end{equation}
where the likelihood function $p\left(\mathbf{y}|A,\mathbf{s},\mathbf{h}\mathrm{,\sigma^{2}}\right)$
is given in (\ref{eq:joint Likelihood-1}), and the term $p\left(A,\mathbf{s},\mathbf{h}\mathrm{,\sigma^{2}}\right)$
stands for the prior information on the unknown quantities. The prior
is assumed to factorize as
\begin{align}
p\left(A,\mathbf{s},\mathbf{h}\mathrm{,\sigma^{2}}\right)= & p\left(A\right)\bigg\{\prod_{n,k,m_{t}}p\left(s_{m_{t}}[n,k]|A\right)\bigg\}\cdot\nonumber \\
 & \cdot\prod_{m_{t},m_{r}}p\left(\boldsymbol{\mathbf{h}}_{m_{t},m_{r}}\right)p\left(\sigma^{2}\right).\label{eq:prior_factorization-1}
\end{align}

\subsubsection{Conventional Prior\label{sub:Conventional-Prior}}

A natural choice for the prior distribution of the unknown variables
$(A,\mathbf{s},\mathbf{h},\sigma^{2})$ is given by $A\sim\mathrm{uniform}\left(\mathcal{A}\right)$,
$s_{m_{t}}[n,k]|A\sim\mathrm{uniform}(A)$, $\boldsymbol{\mathbf{h}}_{m_{t},m_{r}}\sim\mathcal{CN}(\mathbf{0},\alpha\mathbf{I})$,
and $\sigma^{2}\sim\mathcal{IG}\left(\alpha_{0},\beta_{0}\right)$
with fixed parameters $(\alpha,\alpha_{0},\beta_{0})$ \cite{Y. Liu}.
Recall that the inverse Gamma distribution is the conjugate prior
for the Gaussian likelihood at hand, and that uninformative priors
can be obtained by selecting sufficiently large $\alpha$ and $\beta_{0}$
and sufficiently small $\alpha_{0}$ \cite{Robert}. The factorization
(\ref{eq:prior_factorization-1}) implies that, under the prior information,
the variables $A$, $\boldsymbol{\mathbf{h}}_{m_{t},m_{r}}$ and $\sigma^{2}$
are independent of each other, and that the transmitted symbols $s_{m_{t}}[n,k]$
are independent of all the other variables in (\ref{eq:prior_factorization-1})
when conditioned on the modulation $A$. The BN \textsl{$\mathcal{G}_{1}$}
that encodes the factorization given by (\ref{eq:posterior_1}), along
with (\ref{eq:joint Likelihood-1}) and (\ref{eq:prior_factorization-1}),
is shown in Fig. \ref{fig:Dirr_BN-1}. 

\begin{figure}[htbp]
\begin{centering}
\textsf{\includegraphics[width=7.5cm,height=5.5cm]{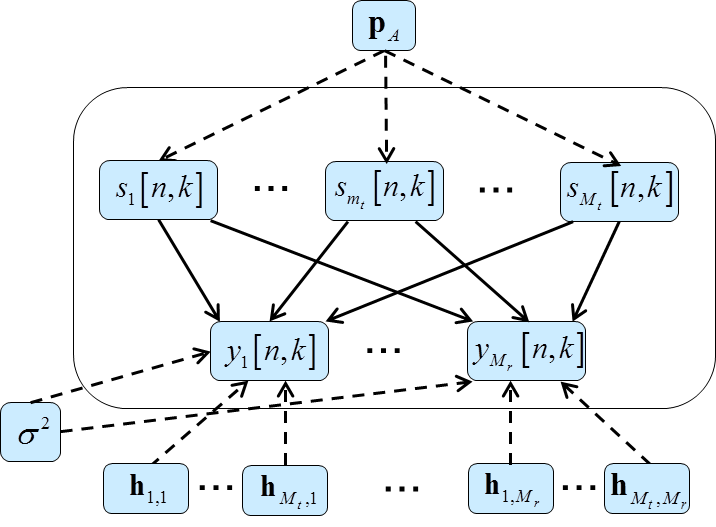}}
\par\end{centering}

\caption{\label{fig:Dirr_BN}BN \textsl{$\mathcal{G}_{2}$} for the modulation
classification scheme based on the Dirichlet latent variable $\mathbf{p}_{A}$.
The nodes inside the rectangle are repeated $NK$ times.}
\end{figure}

The Bayesian inference task for modulation classification of MIMO-OFDM
is to compute the posterior probability of the modulation $A$ conditioned
on the received signal $\mathbf{y}$, namely
\begin{equation}
p\left(A|\mathbf{y}\right)=\sum_{\mathbf{s}}\int p\left(A\text{,}\mathbf{s},\mathbf{h\mathrm{,\sigma^{2}}}|\mathbf{y}\right)d\mathbf{h}d\sigma^{2}.\label{eq:inference_problem-1}
\end{equation}
Following the discussion in Sec. \ref{sec:Preliminaries}, the calculation
in (\ref{eq:inference_problem-1}) is intractable because of the multidimensional
summation and integration. Gibbs sampling (Sec. \ref{sub:Gibbs-sampling})
and mean field variational inference (Sec. \ref{sub:Mean-Field-Variational})
offer feasible alternatives. However, the prior distribution (\ref{eq:prior_factorization-1})
does not satisfy the sufficient condition mentioned in \textit{Remark}
\textit{1}, since some of the conditional distributions required for
Gibbs sampling are not strictly positive in their domains. In particular,
the required conditional distribution for modulation $A$ can be expressed
as
\begin{equation}
p(A=a|\mathbf{s},\mathbf{h\mathrm{,\sigma^{2}}},\mathbf{y})\propto p\left(A\right)\prod_{n,k,m_{t}}p\left(s_{m_{t}}[n,k]|A\right).\label{eq:new}
\end{equation}
The conditional distribution term $p(s_{m_{t}}[n,k]|A=a)$ in (\ref{eq:new})
is zero for all values of $s_{m_{t}}[n,k]$ not belonging to the constellation
$a$, i.e., $p(s_{m_{t}}[n,k]|a)=0$ for $s_{m_{t}}[n,k]\notin a$.
Therefore, the conditional distribution $p(A=a|\mathbf{s},\mathbf{h\mathrm{,\sigma^{2}}},\mathbf{y})$
is equal to zero if the transmitted symbols $\mathbf{s}$ do not belong
to $a$. As a result, the Gibbs sampler may fail to converge to the
posterior distribution (see, e.g., \cite{Amuru}). In order to alleviate
the problem outlined above, we propose to adopt a prior distribution
encoded on a latent Dirichlet BN $\mathcal{G}_{2}$ shown in Fig.
\ref{fig:Dirr_BN}.

\subsubsection{Latent Dirichlet BN\label{sub:Latent-Dirichlet-BN}}

Next, we introduce the Gibbs sampler based on on a latent Dirichlet
BN $\mathcal{G}_{2}$ in details. Accordingly, each transmitted symbol
$s_{m_{t}}[n,k]$ is distributed as a random mixture of uniform distributions
on the different constellations in the set $\mathcal{A}$. Specifically,
a random vector $\mathbf{p}_{A}$ of length $\left|\mathcal{A}\right|$
is introduced to represent the mixture weights, with $\mathbf{p}_{A}(a)$
being the probability that each symbol $s_{m_{t}}[n,k]$ belongs to
the constellation $a\in\mathcal{A}$. Given the mixture weights $\mathbf{p}_{A}$,
the transmitted symbols $s_{m_{t}}[n,k]$ are mutually independent
and distributed according to a mixture of uniform distributions, i.e.,
$p\left(s_{m_{t}}[n,k]|\mathbf{p}_{A}\right)=\sum_{a:\,s_{m_{t}}[n,k]\in a}\mathbf{p}_{A}\left(a\right)/\left|a\right|$.
The Dirichlet distribution is selected as the prior distribution of
$\mathbf{p}_{A}$ in order to simplify the development of the proposed
solutions, as shown in the following subsections. In particular, given
a set of nonnegative parameters $\boldsymbol{\gamma}=[\gamma_{1},\cdots,\gamma_{\left|\mathcal{A}\right|}]^{T}$,
we have $\mathbf{p}_{A}\sim\mathrm{Dirichlet}\left(\boldsymbol{\gamma}\right)$
\cite{Koller}. We recall that the parameter $\gamma_{a}$ has an
intuitive interpretation as it represents the number of symbols in
constellation $a\in\mathcal{A}$ observed during some preliminary
measurements. 

The BN $\mathcal{G}_{2}$ encodes a factorization of the conditional
distribution $p(\mathbf{p}_{A},\mathbf{s},\mathbf{h}\mathrm{,\sigma^{2}}|\mathbf{y})$
\begin{align}
 & p(\mathbf{p}_{A},\mathbf{s},\mathbf{h}\mathrm{,\sigma^{2}}|\mathbf{y})\nonumber \\
\propto & p\left(\mathbf{y}\Big|\mathbf{p}_{A},\mathbf{s},\mathbf{h}\mathrm{,\sigma^{2}}\right)p\left(\mathbf{p}_{A}\right)\bigg\{\prod_{n,k,m_{t}}p\left(s_{m_{t}}[n,k]|\mathbf{p}_{A}\right)\bigg\}\cdot\nonumber \\
 & \cdot\prod_{m_{t},m_{r}}p\left(\boldsymbol{\mathbf{h}}_{m_{t},m_{r}}\right)p\left(\sigma^{2}\right),\label{eq:joint_posterior_2}
\end{align}
where we have $\mathbf{p}_{A}\sim\mathrm{Dirichlet}\left(\boldsymbol{\gamma}\right)$
with a set of nonnegative parameters $\boldsymbol{\gamma}=[\gamma_{1},\cdots,\gamma_{\left|\mathcal{A}\right|}]^{T}$
\cite{Koller}, $p\left(s_{m_{t}}[n,k]|\mathbf{p}_{A}\right)=\sum_{a:\,s_{m_{t}}[n,k]\in a}\mathbf{p}_{A}\left(a\right)/\left|a\right|$,
and the other distributions are as in (\ref{eq:joint Likelihood-1})
and (\ref{eq:prior_factorization-1}). The Bayesian inference task
for modulation classification is to compute the posterior probability
of the mixture weight vector $\mathbf{p}_{A}$ conditional on the
received signal $\mathbf{y}$, namely 
\begin{equation}
p\left(\mathbf{p}_{A}|\mathbf{y}\right)=\sum_{\mathbf{s}}\int p\left(\mathbf{p}_{A}|\mathbf{s},\mathbf{h\mathrm{,\sigma^{2}}}\Big|\mathbf{y}\right)d\mathbf{h}d\sigma^{2},\label{eq:inference_problem}
\end{equation}
and then to estimate $A$ as the value that maximize the a posteriori
mean of $\mathbf{p}_{A}$: 
\begin{equation}
\hat{A}=\arg\max_{a\in\mathcal{A}}\big\langle\mathbf{p}_{A}\left(a\right)\mid\mathbf{y}\big\rangle_{p\left(\mathbf{p}_{A}|\mathbf{y}\right)}.\label{eq:modulation_estimate}
\end{equation}

The proposed approach guarantees that all the conditional distributions
needed for Gibbs sampling based on the BN $\mathcal{G}_{2}$ are non-zero,
and therefore the aforementioned convergence problem for the inference
based on BN $\mathcal{G}_{1}$ is avoided.

\subsection{Modulation Classification via Gibbs Sampling\label{sub:Modulation-Classification-via-1}}

In this subsection, we elaborate on Gibbs sampling for modulation
classification. As explained in Sec. \ref{sub:Gibbs-sampling}, in
order to sample from the joint posterior distribution (\ref{eq:joint_posterior_2}),
the distribution of each variable conditioned on all other variables
is needed. According to (\ref{eq:FullCondionalEq}), we have (for
derivations see Appendix II): 
\begin{enumerate}
\item The conditional distribution of the vector $\mathbf{p}_{A}$, given
$\mathbf{s}$, $\mathbf{h}$, $\mathrm{\sigma^{2}}$ and $\mathbf{y}$
can be expressed as
\begin{equation}
p\left(\mathbf{p}_{A}\Big|\mathbf{s},\mathbf{h}\mathrm{,\sigma^{2}},\mathbf{y}\right)\sim\mathrm{Dirichlet}\left(\boldsymbol{\gamma}+\mathbf{c}\right),\label{eq:P_A posterior}
\end{equation}
where $\mathbf{c}=\left[c_{1},\cdots,c_{\left|\mathcal{A}\right|}\right]^{T}$,
and $c_{a}$ is the number of samples of transmitted symbols in constellation
$a\in\mathcal{A}$; 
\item The distribution of transmitted symbols $s_{m_{t}}[n,k]$, conditioned
on $\mathbf{p}_{A}$, $\mathbf{s}\diagdown s_{m_{t}}[n,k]$, $\mathbf{h}$,
$\mathrm{\sigma^{2}}$ and $\mathbf{y}$, is given by 
\begin{align}
 & p\left(s_{m_{t}}[n,k]\Big|\mathbf{p}_{A},\mathbf{s}\diagdown s_{m_{t}}[n,k],\mathbf{h\mathrm{,\sigma^{2},}}\mathbf{y}\right)\nonumber \\
\propto & p\left(s_{m_{t}}[n,k]|\mathbf{p}_{A}\right)p\left(\mathbf{y}[n,k]\Big|\mathbf{s}[n,k],\mathbf{H}[n],\sigma^{2}\right),\label{eq:tx_symbol_GB}
\end{align}
where we recall that when a new sample is generated for $s_{m_{t}}[n,k]$
, the new value  is updated and used in computing subsequent samples
in $\mathbf{s}$;
\item The required distribution for channel vector $\boldsymbol{\mathbf{h}}_{m_{t},m_{r}}$
is given by 
\begin{align}
 & \boldsymbol{\mathbf{h}}_{m_{t},m_{r}}\Big|\left(\mathbf{P}_{A},\mathbf{s},\mathbf{h}\diagdown\boldsymbol{\mathbf{h}}_{m_{t},m_{r}},\mathbf{\mathrm{\sigma^{2},}}\mathbf{y}\right)\nonumber \\
\sim & \mathcal{CN}(\hat{\mathbf{h}}_{m_{t},m_{r}},\hat{\boldsymbol{\Sigma}}_{m_{t},m_{r}}),\label{eq:channel_GB}
\end{align}
where we have 
\begin{align}
\left(\hat{\boldsymbol{\Sigma}}_{m_{t},m_{r}}\right)^{-1} & =\frac{1}{\mathrm{\sigma^{2}}}\mathbf{W}^{H}\mathbf{D}_{m_{t}}^{H}\left(\mathbf{D}_{m_{t}}\mathbf{W}\right),
\end{align}
and
\begin{align}
\hat{\mathbf{h}}_{m_{t},m_{r}}= & \frac{\hat{\boldsymbol{\Sigma}}_{m_{t},m_{r}}}{\sigma^{2}}\mathbf{W}^{H}\mathbf{D}_{m_{t}}^{H}\cdot\nonumber \\
 & \cdot\bigg(\mathbf{y}_{m_{r}}-\sum_{m_{t}^{\prime}\neq m_{t}}\mathbf{D}_{m_{t}^{\prime}}\tilde{\mathbf{h}}_{m_{t}^{\prime},m_{r}}\bigg);
\end{align}

\item The conditional distribution for $\sigma^{2}$, conditioned on $\mathbf{p}_{A}$,
$\mathbf{s}$, $\mathbf{h}$, and $\mathbf{y}$, is given by
\begin{equation}
\sigma^{2}\Big|\mathbf{p}_{A}\mathbf{s},\mathbf{h},\mathbf{y}\sim\mathcal{IG}\left(\alpha,\beta\right),\label{eq:sigma_sq_GB}
\end{equation}
where $\alpha=\alpha_{0}+NKM_{r}$ and $\beta=\beta_{0}+\sum_{m_{r}}\left\Vert \mathbf{\mathbf{y}}_{m_{r}}-\sum_{m_{t}}\mathbf{D}_{m_{t}}\tilde{\mathbf{h}}_{m_{t},m_{r}}\right\Vert ^{2}$.
Note that (\ref{eq:P_A posterior}) is a consequence of the fact that
Dirichlet distribution is the conjugate prior of the categorical likelihood
\cite{Koller}; (\ref{eq:channel_GB}) can be derived by following
from standard MMSE channel estimation results \cite{Merli}; and (\ref{eq:sigma_sq_GB})
follows the fact that the inverse Gamma distribution is the conjugate
prior for the Gaussian distribution \cite{Mruphy}. 
\end{enumerate}
We summarize the proposed Gibbs sampler for modulation classification
in Algorithm 1.

\begin{algorithm}[h]
\caption{Gibbs Sampling}

\begin{itemize}
\item Initialize $\mathbf{\boldsymbol{\theta}}_{u}^{\left(0\right)}=\{\mathbf{p}_{A}^{(0)},\mathbf{s}^{(0)},\mathbf{h}\mathrm{^{(0)},\sigma^{2}}^{(0)}\}$
from prior distributions as discussed in Sec. \ref{sub:Modulation-Classification-via}
\item \textbf{for} each iteration $m=1:\,M$

\begin{itemize}
\item given $\{\mathbf{s}^{(m-1)},\mathbf{h}\mathrm{^{(m-1)},\sigma^{2}}^{(m-1)}\}$
draw a sample $\mathbf{p}_{A}^{(m)}$ from $p(\mathbf{p}_{A}|\mathbf{s},\mathbf{h}\mathrm{,\sigma^{2}},\mathbf{y})$
in (\ref{eq:P_A posterior});
\item given $\{\mathbf{p}_{A}^{(m)},\mathbf{h}\mathrm{^{(m-1)},\sigma^{2}}^{(m-1)},(\mathbf{s}\diagdown s_{m_{t}}[n,k])^{(m)}\}$,
draw a sample $s_{m_{t}}^{(m)}[n,k]$ from $p(s_{m_{t}}[n,k]|\mathbf{p}_{A},\mathbf{s}\diagdown s_{m_{t}}[n,k],\mathbf{h\mathrm{,\sigma^{2},}}\mathbf{y})$
in (\ref{eq:tx_symbol_GB});
\item given $\{\mathbf{p}_{A}^{(m)},\mathbf{s}\mathrm{^{(m)},\sigma^{2}}^{(m-1)},(\mathbf{h}\diagdown\boldsymbol{\mathbf{h}}_{m_{t},m_{r}})^{(m)}\}$
and the current sample values for , draw a sample $\boldsymbol{\mathbf{h}}_{m_{t},m_{r}}^{(m)}$
from $p(\boldsymbol{\mathbf{h}}_{m_{t},m_{r}}|(\mathbf{P}_{A},\mathbf{s},\mathbf{h}\diagdown\boldsymbol{\mathbf{h}}_{m_{t},m_{r}},\mathbf{\mathrm{\sigma^{2},}}\mathbf{y}))$
in (\ref{eq:channel_GB});
\item given $\{\mathbf{p}_{A}^{(m)},\mathbf{s}^{(m)},\mathbf{h}^{(m)}\}$
draw sample $\mathrm{\sigma^{2}}^{(m)}$ from $p(\sigma^{2}|\mathbf{p}_{A}\mathbf{s},\mathbf{h},\mathbf{y})$
in (\ref{eq:sigma_sq_GB});\textbf{ }
\end{itemize}
\item \textbf{end for}\end{itemize}
\end{algorithm}

\textit{Remark} \textit{3:} In \cite{Thomas}, an alternative Gibbs
sampling approach based on a ``superconstellation'' is proposed
to solve the convergence problem at hand for modulation classification
in SISO. The Gibbs sampling scheme in \cite{Thomas} can be viewed
as an approximation of the approach based on the latent Dirichlet
BN obtained by setting the prior distribution $\mathbf{p}_{A}\sim\mathrm{Dirichlet}\left(\boldsymbol{\gamma}\right)$
such that $\boldsymbol{\gamma}=\mathbf{0}$ and by setting the sample
value of $\mathbf{p}_{A}$ to be equal to the mean of the conditional
distribution $p(\mathbf{p}_{A}|\mathbf{s},\mathbf{h}\mathrm{,\sigma^{2}},\mathbf{y})$,
i.e., $\mathbf{p}_{A}^{(m)}=\mathbf{c}/\sum_{a\in A}c_{a}$ \cite{Y. Liu},
where we recall that $c_{a}$ is the number of symbols that belong
to constellation $a\in\mathcal{A}$. The performance of the ``superconstellation''
approach extended to MIMO OFDM is discussed in Sec. \ref{sec:Numerical-results-and.}.

\textit{Remark 4}: When the SNR is high, the convergence speed is
severely limited by the close-to-zero probabilities in the conditional
distribution (\ref{eq:tx_symbol_GB}). Specifically, as the Gibbs
sampling proceeds with its iterations, the samples of $\sigma^{2}$
tend to be small, making the relationship between $\mathbf{y}[n,k]$
and $s_{m_{t}}[n,k]$ almost deterministic. In particular, the term
$p(\mathbf{y}[n,k]|\mathbf{s}[n,k],\mathbf{H}[n],\sigma^{2})$ in
(\ref{eq:tx_symbol_GB}) satisfies $p(\mathbf{y}[n,k]|\mathbf{s}^{(m)}[n,k],\mathbf{H}^{(m)}[n],(\sigma^{2})^{(m)})\simeq1$
for the selected sample value $\mathbf{s}^{(m)}[n,k]$ , and $p(\mathbf{y}[n,k]|\mathbf{s}^{(m)}[n,k],\mathbf{H}^{(m)}[n],(\sigma^{2})^{(m)})\simeq0$
for all other possible values for $\mathbf{s}[n,k]$. As a result,
transition between states with different values in the Markov chain
occurs with a very low probability leading to extremely slow convergence.
As discussed in \textit{Remark 2}, the strategy of Gibbs sampling
with multiple random restarts and annealing may be adopted to address
this issue. For simulated annealing, we substitute the conditional
distribution (\ref{eq:sigma_sq_GB}) for $\sigma^{2}$ with an iteration
dependent prior given as \cite{Fevotte}
\begin{equation}
\sigma^{2}\Big|\mathbf{p}_{A}\mathbf{s},\mathbf{h},\mathbf{y}\sim\mathcal{IG}\left(\alpha^{\prime},\beta\right),\label{eq:tempering schedule}
\end{equation}
where we have $\alpha^{\prime}(m)=(1-(1-p_{0})\exp(-m/m_{0}))\alpha$,
with $m$ denoting the current iteration index, $p_{0}=0.1$ and $m_{0}=0.3M$,
where $M$ is the total number of iterations. For multiple restarts,
we propose to use the entropy of the pmf $\big\langle\mathbf{p}_{A}\big\rangle_{p\left(\mathbf{p}_{A}|\mathbf{y}\right)}$,
estimated in a run as the metric, to choose among the $N_{run}$ runs
of Gibbs sampling which one should be used in (\ref{eq:modulation_estimate}).
Specifically, the run with the minimum entropy estimate $\big\langle\mathbf{p}_{A}\big\rangle_{p\left(\mathbf{p}_{A}|\mathbf{y}\right)}$
is selected. The rationale of this choice is that an estimate $\big\langle\mathbf{p}_{A}\big\rangle_{p\left(\mathbf{p}_{A}|\mathbf{y}\right)}$
with a low entropy identifies a specific modulation type with a smaller
uncertainty than an estimate $\big\langle\mathbf{p}_{A}\big\rangle_{p\left(\mathbf{p}_{A}|\mathbf{y}\right)}$
with higher entropy (i.e., closer to a uniform distribution).

\subsection{Modulation classification via Mean Field Variational Inference\label{sub:Modulation-classification-via}}

Following the discussion in Sec. \ref{sub:Mean-Field-Variational},
the goal of mean field variational inference applied to the problem
at hand is that of searching for a distribution $q(\mathbf{p}_{A},\mathbf{s},\mathbf{h},\sigma^{2})$
that is closest to the desired posterior distribution $p(\mathbf{p}_{A},\mathbf{s},\mathbf{h},\sigma^{2}|\mathbf{y})$,
in terms of the Kullback-Leibler (KL) divergence $\mathrm{KL}\left(q(\mathbf{p}_{A},\mathbf{s},\mathbf{h},\sigma^{2})||p(\mathbf{p}_{A},\mathbf{s},\mathbf{h},\sigma^{2}|\mathbf{y})\right)$,
within the class $\mathcal{Q}$ of distributions that factorize as
\begin{align}
q(\mathbf{p}_{A},\mathbf{s},\mathbf{h},\sigma^{2}) & =q\left(\mathbf{p}_{A}\right)q\left(\mathbf{s}\right)q\left(\mathbf{h}\right)q\left(\sigma^{2}\right),\label{eq:class of distribution Q}
\end{align}
where $q\left(\mathbf{s}\right)=\prod_{k=1}^{K}\prod_{n=1}^{N}\prod_{m_{t}}^{M_{t}}q(s_{m_{t}}[n,k])$,
and $q\left(\mathbf{h}\right)=\prod_{m_{t}}^{M_{t}}\prod_{m_{r}}^{M_{r}}\boldsymbol{\mathbf{h}}_{m_{t},m_{r}}$.
Next, we present the fixed-point equations for mean field variational
inference. These update equations may be derived by applying (\ref{eq:fixed_point})
to the joint pdf (\ref{eq:joint_posterior_2}). We recall that (\ref{eq:fixed_point})
requires taking expectations of the relevant variables with respect
to updated distribution $q$. If all distributions $q\left(\mathbf{p}_{A}\right)$,
$q\left(\boldsymbol{\mathbf{h}}\right)$ and $q\left(\sigma^{2}\right)$
are initialized by choosing from the conjugate exponential prior family
\cite{Cemgil}, \cite{Ghahramani} in a way being consistent with
the priors in (\ref{eq:joint_posterior_2}), namely $q\left(\mathbf{p}_{A}\right)$
being a Dirichlet distribution, $q\left(\boldsymbol{\mathbf{h}}\right)$
being a circularly complex Gaussian distribution, and $q\left(\sigma^{2}\right)$
being an inverse Gamma distribution, these fixed point update equations
can be calculated, using a similar approach as in the Appendix I,
as follows.
\begin{enumerate}
\item The fixed point update equation for the mixture weight vector $\mathbf{p}_{A}$
can be expressed as 
\begin{equation}
q\left(\mathbf{p}_{A}\right)=\mathrm{Dirichlet}\left(\boldsymbol{\gamma}+\mathbf{g}\right),\label{eq:P_A_MF_2}
\end{equation}
where $\mathbf{g}=\left[g_{1},\cdots,g_{\left|\mathcal{A}\right|}\right]^{T}$
and $g_{a}=\sum_{n,k,m_{t}}\sum_{s_{m_{t}}[n,k]\in a}q(s_{m_{t}}[n,k])$. 
\item The update equation for the transmitted symbol $s_{m_{t}}[n,k]$ is
given by 
\begin{align}
 & q\left(s_{m_{t}}[n,k]\right)\propto\exp\Big[\big\langle\ln p\left(s_{m_{t}}[n,k]|\mathbf{p}_{A}\right)\big\rangle_{q\left(\mathbf{p}_{A}\right)}+\nonumber \\
 & \big\langle\ln p\left(\mathbf{y}[n,k]|\mathbf{s}[n,k],\mathbf{H}[n],\sigma^{2}\right)\big\rangle_{q\left(\mathbf{s}[n,k]\diagdown s_{m_{t}}[n,k],\mathbf{h},\sigma^{2}\right)}\Big]\label{eq:tx_symbol_MF_1}
\end{align}
where the detailed expression of (\ref{eq:tx_symbol_MF_1}) is shown
in Appendix III. 
\item The equation for the channel vector $\boldsymbol{\mathbf{h}}_{m_{t}m_{r}}$
is given by
\begin{equation}
q\left(\boldsymbol{\mathbf{h}}_{m_{t}m_{r}}\right)\mathcal{=CN}(\hat{\mathbf{h}}_{m_{t},m_{r}},\hat{\boldsymbol{\Sigma}}_{m_{t},m_{r}}),\label{eq:channel_MF_2}
\end{equation}
where 
\begin{align}
\left(\hat{\boldsymbol{\Sigma}}_{m_{t},m_{r}}\right)^{-1} & =\frac{\alpha}{\beta}\mathbf{W}^{H}\big\langle\mathbf{D}_{m_{t}}^{H}\mathbf{D}_{m_{t}}\big\rangle\mathbf{W},
\end{align}
 
\begin{align}
 & \hat{\mathbf{h}}_{m_{t},m_{r}}\nonumber \\
= & \frac{\alpha}{\beta}\mathbf{W}^{H}\big\langle\mathbf{D}_{m_{t}}\big\rangle^{H}\bigg(\mathbf{y}_{m_{r}}-\sum_{m_{t}^{\prime}\neq m_{t}}\mathbf{\mathbf{\Lambda}}_{m_{t}^{\prime},m_{r}}\bigg),
\end{align}
\begin{equation}
\mathbf{\mathbf{\Lambda}}_{m_{t}^{\prime},m_{r}}=\big\langle\mathbf{D}_{m_{t}}\big\rangle\mathbf{W}\hat{\mathbf{h}}_{m_{t}^{\prime},m_{r}},
\end{equation}
and $\big\langle\mathbf{D}_{m_{t}}^{H}\mathbf{D}_{m_{t}}\big\rangle$
is a diagonal matrix whose$\left(n,n\right)$ element is equal to
$\sum_{k=1}^{K}\big\langle|s_{m_{t}}[n,k]|^{2}\big\rangle$.
\item The fixed point update equation for the noise variance $\sigma^{2}$
can be expressed as 
\begin{equation}
q\left(\sigma^{2}\right)=\mathcal{IG}\left(\alpha,\beta\right),\label{eq:MF_sig}
\end{equation}
where $\alpha=\alpha_{0}+NKM_{r}$, and $\beta=\beta_{0}+\sum_{m_{r}}\beta_{m_{r}}$
with 
\begin{align}
\beta_{m_{r}}= & \left|\mathbf{y}_{m_{r}}\right|^{2}-2\mathrm{real}\bigg[\mathbf{y}_{m_{r}}^{H}\sum_{m_{t}}\mathbf{\Lambda}_{m_{t},m_{r}}\bigg]+\nonumber \\
 & +\sum_{m_{t}}\mathbf{\Psi}_{m_{t},m_{r}}+\sum_{m_{t}}\mathbf{\Xi}_{m_{t},m_{r}},
\end{align}
\begin{align}
\mathbf{\Psi}_{m_{t},m_{r}}= & \mathrm{tr}\left[\mathbf{W}^{H}\big\langle\mathbf{D}_{m_{t}}^{H}\mathbf{D}_{m_{t}}\big\rangle\mathbf{W}\hat{\boldsymbol{\Sigma}}_{m_{t},m_{r}}\right]+\nonumber \\
 & \hat{\mathbf{h}}_{m_{t},m_{r}}^{H}\mathbf{W}^{H}\big\langle\mathbf{D}_{m_{t}}^{H}\mathbf{D}_{m_{t}}\big\rangle\mathbf{W}\hat{\mathbf{h}}_{m_{t},m_{r}},
\end{align}
and
\begin{equation}
\mathbf{\Xi}_{m_{t},m_{r}}=\hat{\mathbf{h}}_{m_{t},m_{r}}^{H}\mathbf{W}^{H}\big\langle\mathbf{D}_{m_{t}}\big\rangle^{H}\big\langle\mathbf{D}_{m_{t}^{\prime}}\big\rangle\mathbf{W}\hat{\mathbf{h}}_{m_{t}^{\prime},m_{r}},
\end{equation}
where we use $\mathrm{tr}[\cdot]$ to denote the trace of a matrix. 
\end{enumerate}
We summarize the proposed iterative mean field variational inference
for modulation classification in Algorithm 2. 

\begin{algorithm}[h]
\caption{Mean Field Variational Inference}

\begin{itemize}
\item Initialize $q\left(\mathbf{p}_{A}\right)$, $q\left(\mathbf{s}\right)$,
$q\left(\boldsymbol{\mathbf{h}}\right)$ and $q\left(\sigma^{2}\right)$ 
\item \textbf{for} each iteration $m=1:\,M$

\begin{itemize}
\item given the most updated distribution $q\left(\mathbf{s}\right)$, $q\left(\boldsymbol{\mathbf{h}}\right)$
and $q\left(\sigma^{2}\right)$, update the distribution $q\left(\mathbf{p}_{A}\right)$
using (\ref{eq:P_A_MF_2});
\item given the most updated distribution $q\left(\mathbf{p}_{A}\right)$,
$q\left(\mathbf{s}\diagdown s_{m_{t}}[n,k]\right)$, $q\left(\boldsymbol{\mathbf{h}}\right)$
and $q\left(\sigma^{2}\right)$, update the distribution $q\left(s_{m_{t}}[n,k]\right)$
using (\ref{eq:tx_symbol_MF_1});
\item given the most updated distribution $q\left(\mathbf{p}_{A}\right)$,
$q\left(\mathbf{s}\right)$, $q\left(\boldsymbol{\mathbf{h}}\diagdown\boldsymbol{\mathbf{h}}_{m_{t},m_{r}}\right)$
and $q\left(\sigma^{2}\right)$, update the distribution $q\left(\boldsymbol{\mathbf{h}}_{m_{t},m_{r}}\right)$
using (\ref{eq:channel_MF_2});
\item given the most updated distribution $q\left(\mathbf{p}_{A}\right)$,
$q\left(\mathbf{s}\right)$, and $q\left(\boldsymbol{\mathbf{h}}\right)$,
update the distribution $q\left(\sigma^{2}\right)$ according to (\ref{eq:MF_sig});
\end{itemize}
\item \textbf{end for}\end{itemize}
\end{algorithm}

\textit{Remark 5}: While Gibbs sampler is generally known to be superior
to mean field as a method for exploring the global solution space,
the mean field algorithm is known to have better convergence speed
in the vicinity of a local optima \cite{Koller}, \cite{Cemgil},
\cite{Blei}. Following \cite{Cemgil}, we then propose a hybrid strategy
strategy that switches from Gibbs sampling to mean field variational
inference to ``zoom in'' on the local minimum of the optimization
problem (\ref{eq:MF_optimi}). Additional discussion on this point
can be found in the next section. A break-down of the contribution
to the computational complexity of each iteration for the proposed
Gibbs sampler are shown in Table \ref{tab:complexity}. In summary,
the Gibbs sampler requires $\mathcal{O}\{N_{it}M_{t}M_{r}[NK(NL+L^{2}+M_{t}N+M_{t}M_{\mathcal{A}})+L^{3}]\}$
basic arithmetic operations, where $N_{it}$ is the total number of
iterations and $M_{\mathcal{A}}$ is the total number of states of
all possible constellations, i.e., $M_{\mathcal{A}}=\sum_{a\in\mathcal{A}}\left|a\right|$.
At each iteration, the number of the basic arithmetic operations required
for mean field variational inference is of the same order of magnitude
as that for Gibbs sampling. This similarity of the computational complexity
of Gibbs sampling and mean field variational inference is also reported
in \cite{Cemgil}, \cite{Blei} and \cite{Cote}.

\begin{table}[tp]
\caption{\label{tab:complexity}Computational Complexity for Gibbs sampling}

\centering{}%
\begin{tabular}{|c|>{\centering}p{5cm}|}
\hline 
Unknowns & Computational complexity for sampling the variable\tabularnewline
\hline 
\hline 
$\mathbf{p}_{A}$ & $\mathcal{O}\{NK\}$\tabularnewline
\hline 
$\mathbf{s}$ & $\mathcal{O}\{M_{t}^{2}M_{r}NKM_{\mathcal{A}}\}$\tabularnewline
\hline 
$\mathbf{h}$ & $\mathcal{O}\{M_{t}M_{r}[NK(NL+L^{2}+M_{t}N)+L^{3}]\}$\tabularnewline
\hline 
$\sigma^{2}$ & $\mathcal{O}\{M_{t}M_{r}N^{2}K\}$\tabularnewline
\hline 
Total & $ $$\mathcal{O}\{N_{it}M_{t}M_{r}[NK(NL+L^{2}+M_{t}N+M_{t}M_{\mathcal{A}})+L^{3}]\}$\tabularnewline
\hline 
\end{tabular}
\end{table}

\section{Numerical Results and Discussions\label{sec:Numerical-results-and.} }

In this section, we evaluate the performance of the proposed modulation
classification schemes for the detection of three possible modulation
formats within a MIMO-OFDM system. The performance criterion is the
probability of correct classification assuming that all the modulations
are equally likely. Normalized Rayleigh fading channels are assumed
such that $E[\left\Vert \mathbf{h}_{m_{t},m_{r}}\right\Vert ^{2}]=1.$
We define the average SNR as $10\log(M_{t}/\mathrm{\sigma^{2})}$.
Unless stated otherwise, the following conditions are assumed: \textit{i})
$\mathcal{A}$=\{QPSK, 8-PSK, 16QAM\};\textit{ ii}) $M_{t}=M_{r}=2$
antennas; \textit{iii}) $K=2$ OFDM symbols; and \textit{iv}) $L=5$
taps with relative powers given by $[0\mathrm{\,dB},-4.2\mathrm{\,dB},-11.5\,\mathrm{dB},-17.6\mathrm{\,dB},$
$-21.5\,\mathrm{dB}]$.

\subsection{Performance of Gibbs Sampling\label{sub:Performance-of-Gibbs}}

\subsubsection{Gibbs Sampling with Restarts and Annealing\label{sub:Gibbs-Sampling-with}}

We first investigate the performance of the proposed Gibbs sampling
algorithms with or without multiple random restarts and simulated
annealing within each run (see \textit{Remark 4}). The number of runs
in each process of Gibbs sampling with multiple random restarts is
selected to be $N_{run}=5$, and the number of iterations in each
run is $M=2000$, where $M_{0}=0.85M$ initial samples are used as
burn-in period.\footnote{The samples in the burn-in period are not used to evaluate the average
in (\ref{eq:modulation_estimate}).} Note here that the total number of iterations required for Gibbs
sampling is $N_{it}=N_{run}M$. All elements of the vector parameter
$\boldsymbol{\gamma}$ of the prior distribution $\mathbf{p}_{A}\sim\mathrm{Dirichlet}\left(\boldsymbol{\gamma}\right)$
are selected to be equal to a parameter $\gamma$. As also reported
in \cite{Y. Liu}, it may be shown, via numerical results, that the
modulation classification performance is not sensitive to the choice
of parameter $\gamma$ as long as the value of the virtual observation
$\gamma$ (see Sec. \ref{sub:Latent-Dirichlet-BN}) is not very small
($\gamma<1$). For the numerical experiments in this paper, we select
the values of $\gamma$ to be equal to $8\%$ of the total number
of symbols, e.g., in this example $\gamma=[0.08NKM_{t}]=40$. 

\begin{figure}[htbp]
\begin{centering}
\textsf{\includegraphics[width=8.5cm,height=6.05cm]{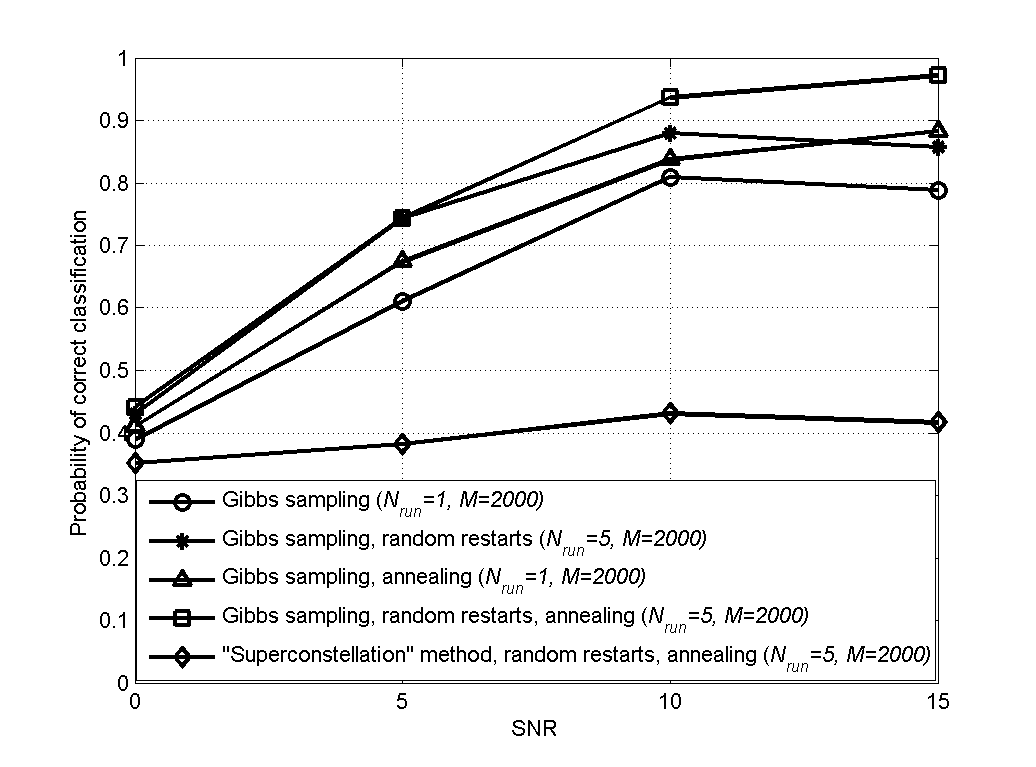}}
\par\end{centering}

\caption{\label{fig:Gibbs}Probability of correct classification using Gibbs
sampling versus SNR ($N=128$, $M_{t}=M_{r}=2$, $K=2$ and $L=5$). }
\end{figure}

In Fig. \ref{fig:Gibbs}, the probabilities of correct classification
for regular Gibbs sampling, Gibbs sampling with multiple random restarts,
Gibbs sampling with annealing and Gibbs sampling with both multiple
random restarts and annealing are plotted as a function of SNR. We
also show for reference the performance of the 'superconstellation'
scheme, with both multiple random restarts and annealing, of \cite{Thomas}
(see \textit{Remark} \textit{3}) extended to MIMO OFDM systems. From
Fig. \ref{fig:Gibbs}, it can be seen that the proposed strategy outperforms
the approach in \cite{Thomas}. Moreover, both strategies of multiple
random restarts and annealing improve the success rate, and that the
best performance is achieved by Gibbs sampling with both random restarts
and annealing. As discussed in \textit{Remark 4}, annealing is seen
to be especially effective in the high-SNR regime. 

\subsubsection{Performance Under Incorrect Channel Length Estimates\label{sub:Performance-Under-Incorrect}}

\begin{figure}[htbp]
\begin{centering}
\textsf{\includegraphics[width=8.5cm,height=6.05cm]{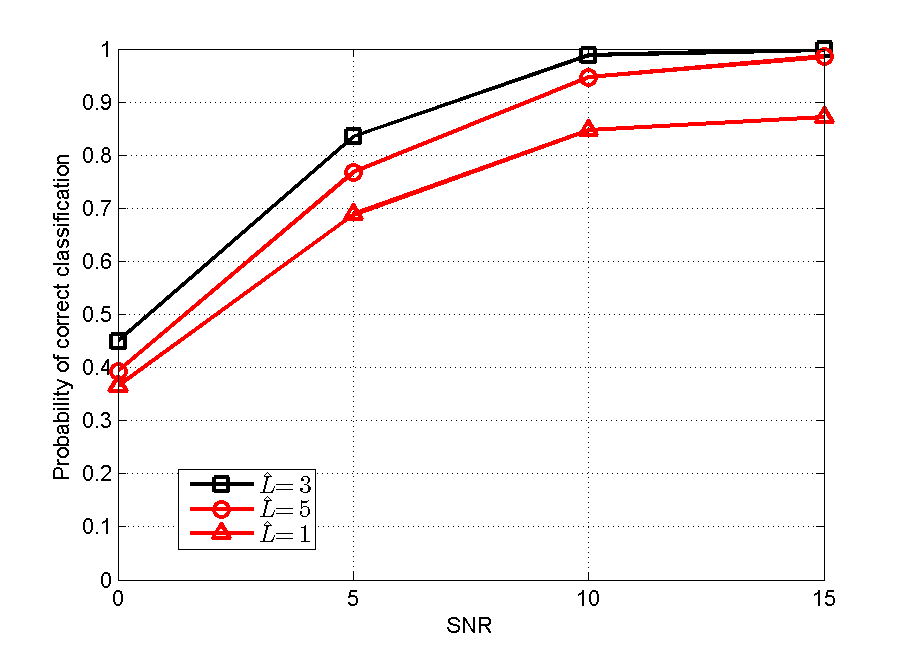}}
\par\end{centering}

\caption{\textcolor{blue}{\label{fig:Gibbs-1}}Probability of correct classification
using Gibbs sampling versus SNR with different channel length estimates
$\hat{L}$ ($N=128$, $M_{t}=M_{r}=2$, $K=2$ and $L=3$). }
\end{figure}

Next, we study the effect of incorrect channel length estimates. The
relative powers of the considered channel taps are $[0\mathrm{\,dB},-2\mathrm{\,dB},-2.5\,\mathrm{dB}]$.
We considered the performance of the proposed scheme under overestimated,
correctly estimated, or underestimated channel lengths. Specifically,
the channel length estimates take three possible values, namely $\hat{L}=1$,
$\hat{L}=3$ or $\hat{L}=5$, while $L=3$. The same values for the
parameters of Gibbs sampler are used as in Sec. \ref{sub:Gibbs-Sampling-with}.
Fig. \ref{fig:Gibbs-1} shows the probabilities of correct classification
for Gibbs sampling with both random restarts and annealing versus
SNR. It is observed that there is a minor performance degradation
with an overestimated channel length. Here, for this example, the
degradation caused by the overfitting when a more complex model with
$\hat{L}=5$ is used is minor. In contrast, a more severe performance
degradation is observed for the case of the underestimated channel
length. This significant degradation is caused by the bias introduced
by the simpler model with $\hat{L}=1$. We also carried out the experiments
for $L=5$ taps with relative powers of $[0\mathrm{\,dB},-2\mathrm{\,dB},-2.5\,\mathrm{dB},$
$-3.1\,\mathrm{dB,}$$-4.2\,\mathrm{dB}]$. The performance is very
similar to that for the case of $L=3$ shown in Fig. \ref{fig:Gibbs-1}
and is hence not reported here. 

\subsubsection{Performance with a Larger $\mathcal{A}$}

To study the effect of modulation pool $\mathcal{A}$ with a larger
size, besides the three modulations considered above, we added 16-PSK
into the modulation pool, i.e., $\mathcal{A}$=\{QPSK, 8-PSK, 16QAM,
16-PSK\}. The relative powers of the multi-path components and the
parameters of the Gibbs sampler take the same value as in Sec. \ref{sub:Performance-Under-Incorrect}.
The performance of the proposed Gibbs sampler for the cases of three
and four modulation formats is shown in Fig. \ref{fig:Gibbs-1}. As
expected, some performance degradation is observed for a larger set
of possible modulation schemes. To gain more insight into the classifier
behavior, the confusion matrices for both cases of three and four
modulation schemes for SNR of 5 dB are shown in Tables \ref{tab:3_modulations}
and \ref{tab:4_modulations}, respectively. The confusion matrices
show the probabilities of deciding for a given modulation format when
another format is the correct one. For instance, the element associated
with row \textquoteleft 8-PSK\textquoteright{} and column \textquoteleft QPSK\textquoteright{}
in Table \ref{tab:3_modulations} indicates that, when the actual
modulation scheme is 8-PSK, the probability that the estimated modulation
is QPSK is 9.2\%. Comparing Table \ref{tab:3_modulations} with Table
\ref{tab:4_modulations}, it can be seen that the decreased accuracy
is mainly caused by the confusion between the two most similar modulation
formats, namely 8-PSK and 16-PSK. 
\begin{figure}[htbp]
\begin{centering}
\textsf{\includegraphics[width=8.5cm,height=6.05cm]{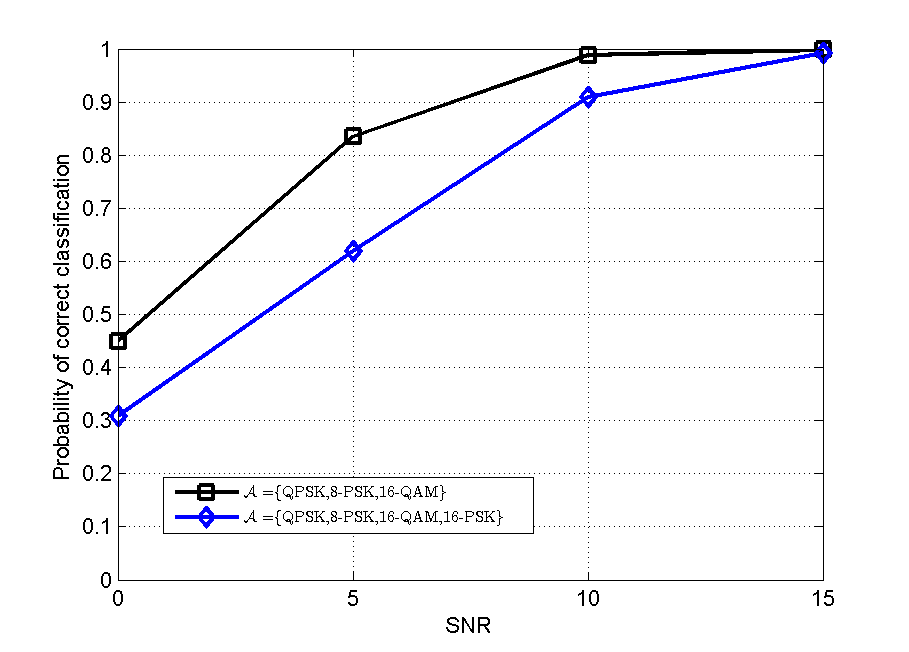}}
\par\end{centering}

\caption{\textcolor{blue}{\label{fig:Gibbs-1-1}}Probability of correct classification
using Gibbs sampling versus SNR with different sets $\mathcal{A}$
of possible modulation schemes ($N=128$, $M_{t}=M_{r}=2$, $K=2$
and $L=3$). }
\end{figure}

\begin{table}[tp]
\caption{\label{tab:3_modulations}Confusion matrix of the proposed Gibbs sampler
for three modulation formats at 5 dB}

\centering{}%
\begin{tabular}{|c|c|c|c|c|}
\cline{3-5} 
\multicolumn{1}{c}{} &  & \multicolumn{3}{c|}{Estimated}\tabularnewline
\cline{3-5} 
\multicolumn{1}{c}{} &  & QPSK & 8-PSK & 16-QAM\tabularnewline
\hline 
\multirow{3}{*}{Actual} & QPSK & 96.3\% & 2.7\% & 1.0\%\tabularnewline
\cline{2-5} 
 & 8-PSK & 9.2\% & 88.1\% & 2.7\%\tabularnewline
\cline{2-5} 
 & 16-QAM & 15.4\% & 18.6\% & 66.0\%\tabularnewline
\hline 
\end{tabular}
\end{table}

\begin{table}[tp]
\caption{\label{tab:4_modulations}Confusion matrix of the proposed Gibbs sampler
for four modulation formats at 5 dB}

\centering{}%
\begin{tabular}{|c|c|c|c|c|c|}
\cline{3-6} 
\multicolumn{1}{c}{} &  & \multicolumn{4}{c|}{Estimated}\tabularnewline
\cline{3-6} 
\multicolumn{1}{c}{} &  & QPSK & 8-PSK & 16-QAM & 16-PSK\tabularnewline
\hline 
\multirow{4}{*}{Actual} & QPSK & 96.8\% & 1.8\% & 0.6\% & 0.8\%\tabularnewline
\cline{2-6} 
 & 8-PSK & 9.4\% & 43.4\% & 3.6\% & 43.6\%\tabularnewline
\cline{2-6} 
 & 16-QAM & 18.8\% & 11.6\% & 60.6\% & 9.0\%\tabularnewline
\cline{2-6} 
 & 16-PSK & 11.8\% & 37.8\% & 3.8\% & 46.6\%\tabularnewline
\hline 
\end{tabular}
\end{table}

\subsection{Performance of Mean Field Variational Inference}

Here, we study the performance of a hybrid scheme, inspired by \cite{Cemgil},
that starts with a Gibbs sampler in order to perform a global search
in the parameter space and then switches to mean field variational
inference to speed up the convergence to a nearby local optima. In
the switching iteration, all the needed marginal distributions for
mean field variational inference are initialized as the conditional
distributions for Gibbs sampling in the previous iteration, namely
the marginal distributions are initialized as follows,
\begin{equation}
q^{(m_{s})}(\mathbf{p}_{A})=p^{(m_{s}-1)}(\mathbf{p}_{A}|\mathbf{s},\mathbf{h}\mathrm{,\sigma^{2}},\mathbf{y}),
\end{equation}
\begin{align}
 & q^{(m_{s})}(s_{m_{t}}[n,k])\nonumber \\
= & p^{(m_{s}-1)}(s_{m_{t}}[n,k]|\mathbf{p}_{A},\mathbf{s}\diagdown s_{m_{t}}[n,k],\mathbf{h\mathrm{,\sigma^{2},}}\mathbf{y}),
\end{align}
\begin{align}
 & q^{(m_{s})}(\boldsymbol{\mathbf{h}}_{m_{t}m_{r}})\nonumber \\
= & p^{(m_{s}-1)}(\boldsymbol{\mathbf{h}}_{m_{t},m_{r}}|\mathbf{P}_{A},\mathbf{s},\mathbf{h}\diagdown\boldsymbol{\mathbf{h}}_{m_{t},m_{r}},\mathbf{\mathrm{\sigma^{2},}}\mathbf{y}),
\end{align}
and
\begin{equation}
q^{(m_{s})}(\sigma^{2})=p^{(m_{s}-1)}(\sigma^{2}|\mathbf{p}_{A}\mathbf{s},\mathbf{h},\mathbf{y}),
\end{equation}
where $m_{s}$ denotes the index of the switching iteration.

\begin{figure}[htbp]
\begin{centering}
\textsf{\includegraphics[width=8.5cm,height=6.05cm]{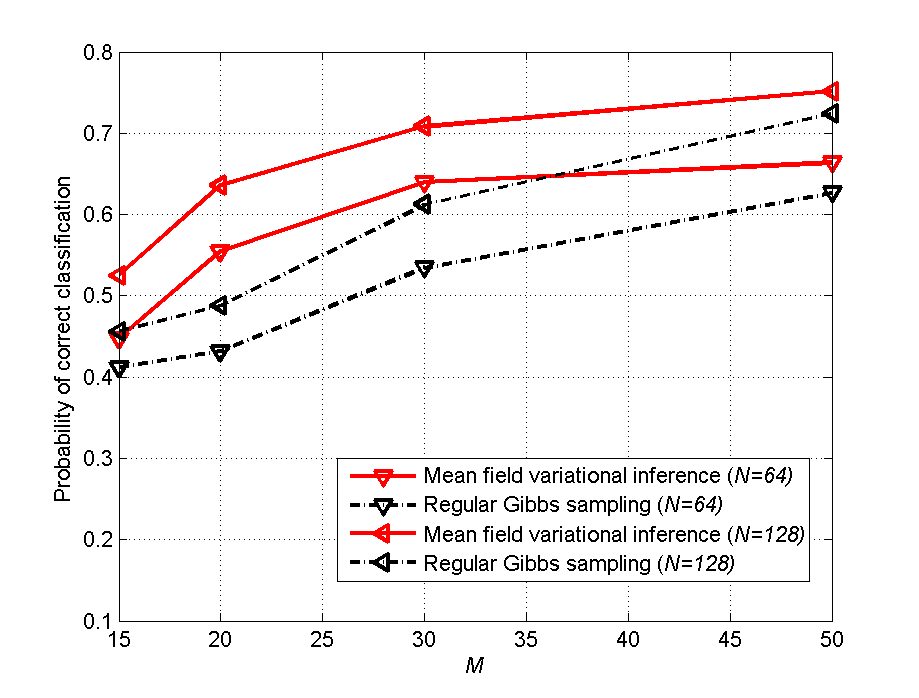}}
\par\end{centering}

\caption{\label{fig:GB=000026MF}Probability of correct classification using
Gibbs sampling and mean field variational inference versus $M$ ($M_{t}=M_{r}=2$,
$K=1$, $L=2$ and SNR=$10\,\mathrm{dB}$). }
\end{figure}

To demonstrate the effectiveness of this approach, we consider modulation
classification using received samples within one OFDM frame ($K=1$)
over Rayleigh fading channels with two taps ($L=2$) and with the
relative powers $[0\mathrm{\,dB},-4.2\,\mathrm{dB}]$. The SNR is
$10\,\mathrm{dB}$ and the DFT size is $N=64$ or $128$. After the
first eight iterations of regular Gibbs sampling (without random restarts
and annealing), we switch to the mean field variational inference.
The probability of correct classification of both regular Gibbs sampling
and the hybrid approach versus the number of iterations $M$ with
$M_{0}=0.9M$ burn-in samples is shown in Fig. \ref{fig:GB=000026MF}.
It can be observed that the hybrid approach outperforms Gibbs sampling
especially when the number of iterations is small.

\subsection{Comparison of Gibbs Sampling and ICA-PC \cite{Handan}}

Here, we compare the classification results achieved by the proposed
Gibbs sampling scheme with the ICA-PC approach of \cite{Handan},
which extends to MIMO-OFDM the techniques studied in \cite{Choqueuse}.
The approach in \cite{Handan} exploits the invariance of the frequency-domain
channels across the coherence bandwidth to perform classification.
Specifically, the subcarriers are grouped in sets of $D$ adjacent
subcarriers whose frequency-domain channel matrices are assumed to
be identical. Let us denote the frequency-domain channel matrix and
the received samples for the $i$-th group by $\mathbf{H}_{i}$ and
$\mathbf{y}_{i}$ respectively, $i=1,...,N/D$. To compute the likelihood
function $\,p(\mathbf{y}_{i}|A=a,\mathbf{H}_{i})$ of the received
samples $\mathbf{y}_{i}$ over the subcarriers within group $i$,
an estimate $\mathbf{\hat{H}}_{i}$ of the channel matrix $\mathbf{H}_{i}$
is first obtained using ICA-PC, and then the likelihood $p(\mathbf{y}_{i}|A=a,\mathbf{H}_{i})$
is approximated as $p(\mathbf{y}_{i}|A=a,\mathbf{\hat{H}}_{i})$.
Accordingly, the likelihood function $p(\mathbf{y}|A=a,\mathbf{H})$
of all the received samples $\mathbf{y}$ is approximated as $p(\mathbf{y}|A=a,\mathbf{\hat{H}})=\prod_{i}p(\mathbf{y}_{i}|A=a,\mathbf{\hat{H}}_{i})$,
where $\mathbf{\widehat{H}}=\{\mathbf{\hat{H}}_{i}\}_{i=1}^{N/D}$.
The detected modulation is selected as $\widehat{A}=\arg\max_{a\in\mathcal{A}}p(\mathbf{y}|A=a,\mathbf{\hat{H}})$. 

\begin{figure}[htbp]
\begin{centering}
\textsf{.\includegraphics[width=8.5cm,height=6.2cm]{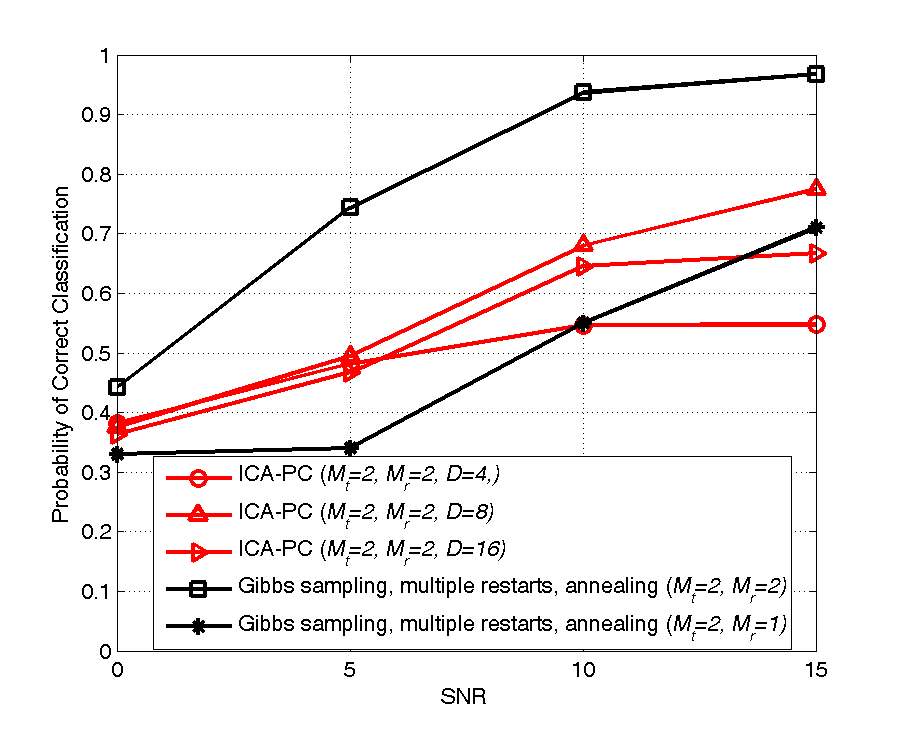}}
\par\end{centering}

\caption{\label{fig:ICA=000026GIBBS}Probability of correct classification
using Gibbs sampling with multiple random restarts and annealing and
approach of \cite{Handan} based on ICA-PC versus SNR ($N=128$, $K=2$
and $L=5$). }
\end{figure}

In Fig. \ref{fig:ICA=000026GIBBS}, we plot the performance of the
approach based on ICA-PC with different values of $D$ and Gibbs sampling
with random restarts and annealing. The number of runs are $N_{run}=5$,
and the annealing schedule is (\ref{eq:tempering schedule}). It can
be seen from Fig. \ref{fig:ICA=000026GIBBS} that Gibbs sampling significantly
outperforms ICA-PC. In this regard, note that, with $D=4$, the accuracy
in ICA-PC is poor due to the insufficient number of observed data
samples; while with $D=16$, the model mismatch problem becomes more
severe due to the assumption of equal channel matrices in each subcarrier
group. 

To further emphasize the advantages of the proposed Bayesian techniques,
in Fig. \ref{fig:ICA=000026GIBBS}, we consider the case in which
there are fewer receive antennas than transmit antennas by setting
$M_{r}=1$ and all other parameters as above. While ICA-PC cannot
be applied due to the ill-posedness of the ICA problem, it can be
observed that a success rate of $71\%$ can be attained by the proposed
Gibbs scheme at $15\mathrm{\,dB}$. It should be mentioned however,
that the complexity of ICA-PC of the order of $\mathcal{O}\{M_{t}M_{r}KNM_{u}^{M_{t}}\}$
\cite{Muhlhaus-1} is smaller than that of Gibbs sampling (see \textit{Remark
5}), where $M_{u}$ denotes the maximum number of states of a constellation
among all the possible modulation types, i.e., $M_{u}=\max_{a\in\mathcal{A}}\{\left|a\right|\}$. 

\subsection{Performance for Coded OFDM Signals}

\begin{figure}[htbp]
\begin{centering}
\textsf{\includegraphics[width=8.5cm,height=6.05cm]{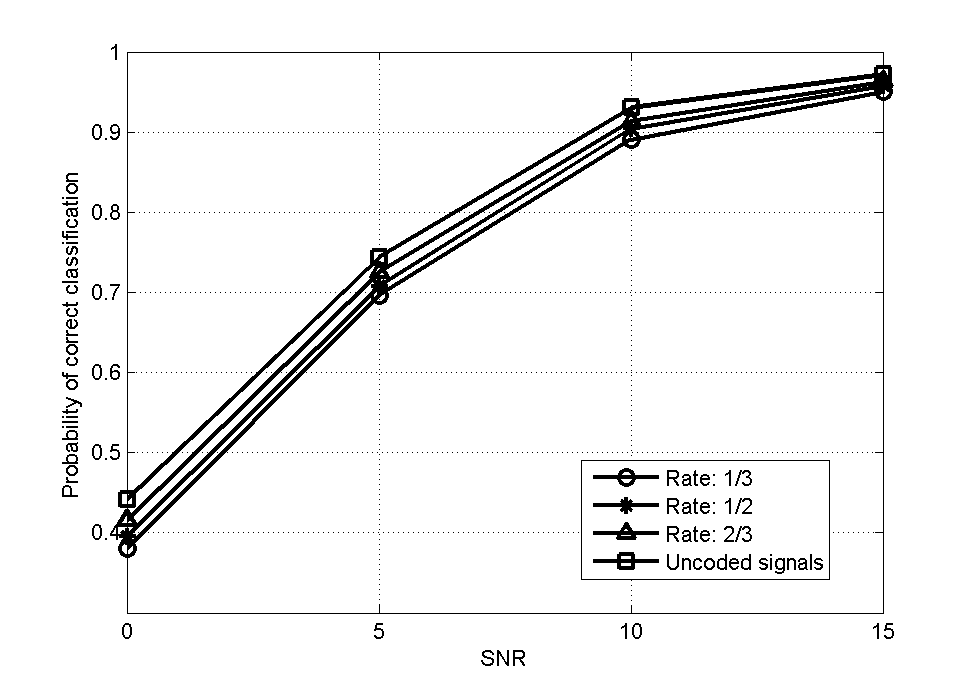}}
\par\end{centering}

\caption{\label{fig:codedOFDM}Probability of correct classification using
Gibbs sampling with multiple random restarts and annealing versus
SNR for convolution coded OFDM signals with different code rate ($N=128$,
$M_{t}=M_{r}=2$, $K=2$ and $L=5$). }
\end{figure}

To investigate the performance of the proposed Gibbs sampling approach
in the presence of a model mismatch, we study here the case of coded
OFDM. Specifically, we assume that the information bits are first
encoded using a convolutional code, and then modulated. We apply Gibbs
sampling with multiple random restarts and annealing with all relevant
parameters being the same as in Sec. \ref{sub:Performance-of-Gibbs}.
The code rates are $1/3,$ $1/2$ and $2/3$, respectively. In Fig.
\ref{fig:codedOFDM}, the probability of correct classification is
shown versus SNR. It can be seen from the figure that the success
rate decreases only slightly (up to 6\%) as the code rate decreases.
The degradation is caused by the fact that the coded transmitted symbols
are not mutually independent due to the convolutional coding, and
hence their prior distribution is mismatched with respect to the model
(\ref{eq:joint_posterior_2}). As the SNR increases, the performance
degradation for coded signals become minor, because, in this regime,
Gibbs sampling relies more on the observed samples than on the priors.
We also studied the performance of the proposed Gibbs sampling for
OFDM systems with space-frequency coded symbols (SF-OFDM) \cite{Lee}.
Compared to the uncoded case, minor performance degradation (2\% at
$15\mathrm{\,dB}$ and up to 8\% at $0\mathrm{\,dB}$) is observed
also due to the presence of a model mismatch. 

\section{Conclusions\label{sec:Conclusions}}

In this paper, we have proposed two Bayesian modulation classification
schemes for MIMO-OFDM systems based on a selection of the prior distributions
that adopts a latent Dirichlet model and on the Bayesian network formalism.
The proposed Gibbs sampling method converges to an effective solution
and, using numerical results, its accuracy is demonstrated to improve
for small sample sizes when switching to the mean field variational
inference technique after a number of iterations. The speed of convergence
is shown to improve via multiple random restarts and annealing. The
techniques are seen to overcome the performance limitation of state-of-the-art
non-Bayesian schemes based on ICA and Bayesian schemes based on ``superconstellation''
methods. In fact, while most of the mentioned existing modulation
classification algorithms rely on the assumptions that the channels
are flat fading, that the number of receive antennas is no less than
the number of transmit antenna, and/or that a large amount of samples
are available (as for pattern recognition-based methods), the proposed
schemes achieve satisfactory performance under more general conditions.
For example, with $M_{t}=2$ transmit antennas and under frequency
selective fading channels with $L=5$ taps, a correct classification
rate of above $97\%$ may be attained with $M_{r}=2$ receive antennas
and with 256 received samples at each antenna; and a success rate
of above $70\%$ may be achieved with $M_{r}=1$ receive antenna and
256 received samples at the antenna. Moreover, the proposed Gibbs
sampler presents a graceful degradation in the presence of a model
mismatch caused by channel coding, e.g., a decrease in the success
rate by $6\%$ with a code rate of $1/3$. Future works include devising
a Gibbs sampling scheme that accounts for the effects of the timing
and carrier frequency offsets for MIMO systems following, e.g., \cite{Amuru},
\cite{Merli}, \cite{Lu}. In addition, the development of Bayesian
classification techniques that address non-Gaussian noise is also
a topic for further investigation.

\section*{Appendix I}

\section*{Distributions}

In this Appendix, we give the expressions for all standard distributions
that are useful to derive the conditional distributions for Gibbs
sampling in Appendix II. 
\begin{enumerate}
\item Dirichlet Distribution: $\mathbf{Z}\sim\mathrm{Dirichlet}\left(\mathbf{c}\right),$
\begin{equation}
p\left(\mathbf{Z}\right)=\frac{\Gamma\left(\sum_{i=1}^{k_{z}}c_{i}\right)}{\prod_{i=1}^{k_{z}}\Gamma\left(c_{i}\right)}\prod_{i=1}^{k_{z}}z_{i}^{c_{i}-1},
\end{equation}
where $k_{z}$ denotes the length of the vector $\mathbf{Z}$, and
$\Gamma(\cdot)$ stands for the gamma function \cite{Abramowitz}.
\item Circular complex Gaussian distribution: $\mathbf{Z}\sim\mathcal{CN}(\mathbf{\mu},\mathbf{\Sigma})$,
\begin{equation}
p\left(\mathbf{Z}\right)=\frac{1}{\pi^{k_{z}}\det\left(\mathbf{\Sigma}\right)}\exp\left\{ -\left(\mathbf{Z}-\mathbf{\mu}\right)^{H}\mathbf{\Sigma}^{-1}\left(\mathbf{Z}-\mathbf{\mu}\right)\right\} ,
\end{equation}
where we use $\det(\cdot)$ to denote the determinant of a matrix.
\item Inverse Gamma distribution:: $z\sim\mathcal{IG}(c,d)$,
\begin{equation}
p\left(\mathbf{Z}\right)=\frac{d^{c}}{\Gamma\left(c\right)}z^{-c-1}\exp\left(-\frac{d}{z}\right).
\end{equation}

\end{enumerate}

\section*{Appendix II}

\section*{Derivations of Conditional Distributions for Gibbs Sampling}

In this Appendix, the required conditional distributions for Gibbs
sampling are derived.

\subsection*{A. Expression for $p(\mathbf{p}_{A}|\mathbf{s},\mathbf{h}\mathrm{,\sigma^{2}},\mathbf{y})$}

Applying (\ref{eq:FullCondionalEq}) to (\ref{eq:joint_posterior_2}),
we have
\begin{align}
 & p\left(\mathbf{p}_{A}\Big|\mathbf{s},\mathbf{h}\mathrm{,\sigma^{2}},\mathbf{y}\right)\nonumber \\
\propto & p\left(\mathbf{p}_{A}\right)\bigg\{\prod_{n,k,m_{t}}p\left(s_{m_{t}}[n,k]|\mathbf{p}_{A}\right)\bigg\}\nonumber \\
\propto & \prod_{a\in\mathcal{A}}\left[\mathbf{p}_{A}\left(a\right)\right]^{\boldsymbol{\gamma}\left(a\right)-1}\prod_{n,k,m_{t}}\bigg[\sum_{a:\,s_{m_{t}}[n,k]\in a}\mathbf{p}_{A}\left(a\right)/\left|a\right|\bigg]\nonumber \\
= & \prod_{a\in\mathcal{A}}\frac{1}{\left|a\right|}\left[\mathbf{p}_{A}\left(a\right)\right]^{\boldsymbol{\gamma}\left(a\right)-1+c_{a}}\nonumber \\
\sim & \mathrm{Dirichlet}\left(\boldsymbol{\gamma}+\mathbf{c}\right),
\end{align}
where $\mathbf{c}=\left[c_{1},\cdots,c_{\left|\mathcal{A}\right|}\right]^{T}$,
and $c_{a}$ is the number of samples of transmitted symbols in constellation
$a\in\mathcal{A}$.

\subsection*{B. Expression for $p(\boldsymbol{\mathbf{h}}_{m_{t},m_{r}}|(\mathbf{P}_{A},\mathbf{s},\mathbf{h}\diagdown\boldsymbol{\mathbf{h}}_{m_{t},m_{r}},\mathbf{\mathrm{\sigma^{2},}}\mathbf{y}))$}

Applying (\ref{eq:FullCondionalEq}) to (\ref{eq:joint_posterior_2}),
we have
\begin{align}
 & p\left(\boldsymbol{\mathbf{h}}_{m_{t},m_{r}}\Big|\left(\mathbf{P}_{A},\mathbf{s},\mathbf{h}\diagdown\boldsymbol{\mathbf{h}}_{m_{t},m_{r}},\mathbf{\mathrm{\sigma^{2},}}\mathbf{y}\right)\right)\nonumber \\
\propto & p\left(\boldsymbol{\mathbf{h}}_{m_{t},m_{r}}\right)p\left(\mathbf{y}\Big|\mathbf{p}_{A},\mathbf{s},\mathbf{h}\mathrm{,\sigma^{2}}\right)\nonumber \\
\propto & \exp\bigg\{-\boldsymbol{\mathbf{h}}_{m_{t},m_{r}}^{H}\left(\hat{\boldsymbol{\Sigma}}_{m_{t},m_{r}}\right)^{-1}\boldsymbol{\mathbf{h}}_{m_{t},m_{r}}+\nonumber \\
 & 2\mathrm{real}\left[\boldsymbol{\mathbf{h}}_{m_{t},m_{r}}^{H}\left(\hat{\boldsymbol{\Sigma}}_{m_{t},m_{r}}\right)^{-1}\hat{\mathbf{h}}_{m_{t},m_{r}}\boldsymbol{\mathbf{h}}_{m_{t},m_{r}}\right]\bigg\}\nonumber \\
\sim & \mathcal{CN}(\hat{\mathbf{h}}_{m_{t},m_{r}},\hat{\boldsymbol{\Sigma}}_{m_{t},m_{r}}),
\end{align}
where
\begin{align}
\left(\hat{\boldsymbol{\Sigma}}_{m_{t},m_{r}}\right)^{-1} & =\frac{1}{\alpha_{h}}\mathbf{I}+\frac{1}{\mathrm{\sigma^{2}}}\mathbf{W}^{H}\mathbf{D}_{m_{t}}^{H}\mathbf{D}_{m_{t}}\mathbf{W}\nonumber \\
 & \approx\frac{1}{\mathrm{\sigma^{2}}}\mathbf{W}^{H}\mathbf{D}_{m_{t}}^{H}\mathbf{D}_{m_{t}}\mathbf{W},\label{eq:h_sig}
\end{align}
and
\begin{align}
\hat{\mathbf{h}}_{m_{t},m_{r}}= & \hat{\boldsymbol{\Sigma}}_{m_{t},m_{r}}\frac{1}{\sigma^{2}}\mathbf{W}^{H}\mathbf{D}_{m_{t}}^{H}\cdot\nonumber \\
 & \cdot\bigg(\mathbf{y}_{m_{r}}-\sum_{m_{t}^{\prime}\neq m_{t}}\mathbf{D}_{m_{t}^{\prime}}\tilde{\mathbf{h}}_{m_{t}^{\prime},m_{r}}\bigg),\label{eq:h_mean}
\end{align}
where the approximation in (\ref{eq:h_sig}) follows from the fact
that $\alpha_{h}$ is very large such that the term $1/\alpha_{h}\mathbf{I}$
can be neglected.

\subsection*{C. Expression for $p\left(\sigma^{2}|\mathbf{p}_{A}\mathbf{s},\mathbf{h},\mathbf{y}\right)$}

Applying (\ref{eq:FullCondionalEq}) to (\ref{eq:joint_posterior_2}),
we have
\begin{align}
 & p\left(\sigma^{2}\Big|\mathbf{p}_{A}\mathbf{s},\mathbf{h},\mathbf{y}\right)\nonumber \\
\propto & p\left(\sigma^{2}\right)p\left(\mathbf{y}\Big|\mathbf{p}_{A},\mathbf{s},\mathbf{h}\mathrm{,\sigma^{2}}\right)\nonumber \\
\propto & \left(\sigma^{2}\right)^{-\alpha_{0}-1}\exp\left(-\frac{\beta_{0}}{\sigma^{2}}\right)\nonumber \\
 & \prod_{m_{r=1}}^{M_{r}}\frac{1}{\sigma^{2NK}}\exp\bigg\{-\frac{1}{\mathrm{\sigma^{2}}}\left\Vert \mathbf{y}_{m_{r}}-\sum_{m_{t=1}}^{M_{t}}\mathbf{D}_{m_{t}}\mathbf{\mathbf{W}}\mathbf{h}_{m_{t},m_{r}}\right\Vert ^{2}\bigg\}\nonumber \\
\propto & \left(\mathrm{\sigma^{2}}\right)^{-(\alpha_{0}+M_{r}NK)-1}\cdot\nonumber \\
 & \cdot\exp\bigg(-\frac{\beta_{0}+\sum_{m_{r=1}}^{M_{r}}\left\Vert \mathbf{y}_{m_{r}}-\sum_{m_{t=1}}^{M_{t}}\mathbf{D}_{m_{t}}\mathbf{\mathbf{W}}\mathbf{h}_{m_{t},m_{r}}\right\Vert ^{2}}{\sigma^{2}}\bigg)\nonumber \\
\sim & \mathcal{IG}\left(\alpha,\beta\right),
\end{align}
where $\alpha=\alpha_{0}+NKM_{r}$ and $\beta=\beta_{0}+\sum_{m_{r}}\left\Vert \mathbf{\mathbf{y}}_{m_{r}}-\sum_{m_{t}}\mathbf{D}_{m_{t}}\tilde{\mathbf{h}}_{m_{t},m_{r}}\right\Vert ^{2}$.

\section*{Appendix III }

\section*{Evaluation of (\ref{eq:tx_symbol_MF_1})}

By taking expectation with respect to updated distribution $q\left(\mathbf{\mathbf{\mathbf{p}_{A}}}\right)$
and $q\left(\mathbf{s}[n,k]\diagdown s_{m_{t}}[n,k],\mathbf{h},\sigma^{2}\right)$
receptively, it can be shown that the expression for $\big\langle\ln p\left(s_{m_{t}}[n,k]|\mathbf{p}_{A}\right)\big\rangle_{\mathbf{p}_{A}}$
is 
\begin{align}
 & \big\langle\ln p\left(s_{m_{t}}[n,k]|\mathbf{p}_{A}\right)\big\rangle_{q\left(\mathbf{\mathbf{\mathbf{p}_{A}}}\right)}\nonumber \\
= & \sum_{a\in\mathcal{A}}\mathrm{\mathbf{1}}\left(s_{m_{t}}[n,k]\in a\right)\left[\psi\left(\gamma_{a}\right)-\psi\left(\gamma_{0}\right)-\ln\left|a\right|\right],
\end{align}
where $\gamma_{0}=\sum_{a\in\mathcal{A}}\gamma_{a}$; and the expression
for $\big\langle\ln p\left(\mathbf{y}[n,k]|\mathbf{s}[n,k],\mathbf{H}[n],\sigma^{2}\right)\big\rangle$
is 
\begin{align}
 & \big\langle\ln p\left(\mathbf{y}[n,k]|\mathbf{s}[n,k],\mathbf{H}[n],\sigma^{2}\right)\big\rangle_{q\left(\mathbf{s}[n,k]\diagdown s_{m_{t}}[n,k],\mathbf{h},\sigma^{2}\right)}\nonumber \\
\propto & \frac{\alpha}{\beta}\bigg\{2\mathrm{real}\left[\mathbf{y}^{H}[n,k]\right]\big\langle\mathbf{H}[n]\big\rangle\big\langle\mathbf{s}[n,k]\big\rangle_{q\left(\mathbf{s}[n,k]\diagdown s_{m_{t}}[n,k]\right)}\nonumber \\
 & -\mathrm{tr}\left(\big\langle\mathbf{H}^{H}[n]\mathbf{H}[n]\big\rangle\mathrm{cov\left(\mathbf{s}[n,k]\right)}\right)\nonumber \\
 & -\big\langle\mathbf{s}[n,k]\big\rangle_{q\left(\mathbf{s}[n,k]\diagdown s_{m_{t}}[n,k]\right)}^{H}\cdot\nonumber \\
 & \cdot\big\langle\mathbf{H}^{H}[n]\mathbf{H}[n]\big\rangle\big\langle\mathbf{s}[n,k]\big\rangle_{q\left(\mathbf{s}[n,k]\diagdown s_{m_{t}}[n,k]\right)}\bigg\},
\end{align}
with the $(m_{t}^{\prime},m_{t}^{\prime\prime})$ element of the covariance
matrix of the vector $\mathbf{s}[n,k]$ 
\begin{align}
 & \mathrm{cov}\left(\mathbf{s}[n,k]\right)_{(m_{t}^{\prime},m_{t}^{\prime\prime})}\nonumber \\
= & \mathrm{var(}s_{m_{t}^{\prime}}[n,k])\nonumber \\
= & \begin{cases}
\big\langle\left|s_{m_{t}^{\prime}}[n,k]\right|^{2}\big\rangle-\big\langle\left|s_{m_{t}^{\prime}}[n,k]\right|\big\rangle^{2}, & \mathrm{if}\:m_{t}^{\prime}=m_{t}^{\prime\prime},\\
 & m_{t}^{\prime\prime}\neq m_{t};\\
0, & \mathrm{otherwise}.
\end{cases}
\end{align}
 the $m_{t}^{\prime}$-th element of $\big\langle\mathbf{s}[n,k]\big\rangle_{q\left(\mathbf{s}[n,k]\diagdown s_{m_{t}}[n,k]\right)}$
being 
\begin{align}
\big\langle\mathbf{s}[n,k]\big\rangle_{(m_{t}^{,})} & =\begin{cases}
\big\langle s_{m_{t}^{\prime}}[n,k]\big\rangle, & \mathrm{if\:}m_{t}^{\prime}\neq m_{t}\\
s_{m_{t}}[n,k], & \mathrm{if}\:m_{t}^{\prime}=m_{t}
\end{cases},
\end{align}
 the $(m_{r}^{\prime},m_{t}^{\prime})$ element of $\big\langle\mathbf{H}[n]\big\rangle$
being the $n$-th element of the matrix product $\mathbf{W\mathbf{\widehat{h}}_{m_{t}^{\prime},m_{r}^{\prime}}}$,
and the $(m_{t}^{\prime},m_{t}^{\prime\prime})$ element of $\big\langle\mathbf{H}^{H}[n]\mathbf{H}[n]\big\rangle$
being 
\begin{align}
 & \left[\big\langle\mathbf{H}^{H}[n]\mathbf{H}[n]\big\rangle\right]_{\left(m_{t}^{\prime},m_{t}^{\prime\prime}\right)}=\nonumber \\
 & \begin{cases}
\sum_{m_{r}=1}^{M_{r}}\Bigg\{\mathrm{tr}\left[\left[\mathbf{W}_{(n,\cdot)}\right]^{H}\mathbf{W}_{(n,\cdot)}\mathbf{\boldsymbol{\Sigma}}_{m_{t}^{\prime\prime},m_{r}}\right]+\\
\big\langle\mathbf{h}_{m_{t}^{\prime\prime},m_{r}}\big\rangle^{H}\left[\mathbf{W}_{(n,\cdot)}\right]^{H}\mathbf{W}_{(n,\cdot)}\big\langle\mathbf{h}_{m_{t}^{\prime\prime},m_{r}}\big\rangle\Bigg\}, & \mathrm{if\:}m_{t}^{\prime}=m_{t}^{\prime\prime};\\
\big\langle\mathbf{H}[n]\big\rangle_{\left(\cdot,m_{t}^{\prime}\right)}\big\langle\mathbf{H}[n]\big\rangle_{\left(\cdot,m_{t}^{\prime\prime}\right)}, & \mathrm{if}\:m_{t}^{\prime}\neq m_{t}^{\prime\prime}.
\end{cases}
\end{align}

\end{document}